\documentclass[12pt,preprint]{aastex}

\shorttitle{Low-$l$ solar $p$ modes and solar abundances}

\shortauthors{Chaplin et al.}

\begin{document}

\title{Solar heavy element abundance: constraints from frequency
separation ratios of low-degree $p$ modes}

\author{William J. Chaplin}

\affil{School of Physics and Astronomy, University of Birmingham,
Edgbaston, Birmingham B15 2TT, U.K.; w.j.chaplin@bham.ac.uk}

\author{Aldo~M.~Serenelli}

\affil{Institute for Advanced Study, Einstein Drive, Princeton, NJ
08540; aldos@ias.edu}

\author{Sarbani Basu}

\affil{Department of Astronomy, Yale University, P.O. Box 208101, New
Haven, CT 06520-8101; sarbani.basu@yale.edu}

\author{Yvonne~Elsworth}

\affil{School of Physics and Astronomy, University of Birmingham,
Edgbaston, Birmingham B15 2TT, U.K.; y.p.elsworth@bham.ac.uk}

\author{Roger~New}

\affil{Faculty of Arts, Computing, Engineering and Sciences, Sheffield
Hallam University, Sheffield S1 1WB, U.K.; r.new@shu.ac.uk}

\and

\author{Graham~A.~Verner}

\affil{School of Physics and Astronomy, University of Birmingham,
Edgbaston, Birmingham B15 2TT, U.K.; gav@bison.ph.bham.ac.uk}

\begin{abstract}

We use very precise frequencies of low-degree solar-oscillation modes
measured from 4752 days of data collected by the Birmingham
Solar-Oscillations Network (BiSON) to derive seismic information on
the solar core.  We compare these observations to results from a large
Monte Carlo simulation of standard solar models, and use the results
to constrain the mean molecular weight of the solar core, and the
metallicity of the solar convection zone.  We find that only a high
value of solar metallicity is consistent with the seismic
observations.  We can determine the mean molecular weight of the solar
core to a very high precision, and, dependent on the sequence of Monte
Carlo models used, find that the average mean molecular weight in the
inner 20\% by radius of the Sun ranges from 0.7209 to 0.7231, with
uncertainties of less than 0.5\% on each value.  Our lowest seismic
estimate of solar metallicity is $Z=0.0187$ and our highest is
$Z=0.0239$, with uncertainties in the range of 12--19\%.  Our results
indicate that the discrepancies between solar models constructed with
low metallicity and the helioseismic observations extend to the solar
core and thus cannot be attributed to deficiencies in the modeling of
the solar convection zone.

\end{abstract}


\section{Introduction}
\label{sec:intro}

In a series of papers, Asplund et al.  (2004, 2005b, 2005c) and
Allende-Prieto et al. (2001, 2002) have revised the spectroscopic
determinations of the solar photospheric composition. In particular,
their results have determined carbon, nitrogen and oxygen abundances
to be lower by about 25\% to 35\% than previous determinations
(Grevesse \& Sauval 1998; hereafter GS98). The revision of the oxygen
abundance leads to a comparable change in the abundances of neon and
argon. Additionally, Asplund (2000) has also determined a somewhat
lower value (10\%) for the photospheric abundance of silicon compared
to the GS98 value. As a result, all the elements for which abundances
are obtained from meteoritic measurements have seen their abundances
reduced by a similar amount. These measurements have been summarized
in Asplund et al. (2005a; hereafter AGS05), and the net result is that
the ratio of the mass fraction of heavy elements to hydrogen in the
Sun is $Z/X=0.0165$ (alternatively, $Z=0.0122$), about 28\% lower than
the previous value, $Z/X=0.0229$ ($Z=0.0169$) given by GS98.

The new low-abundance value for the heavy elements, albeit the result
of a much more sophisticated modeling of the solar atmosphere, has
given rise to discrepancies between helioseismic observations and
predictions from solar models constructed with the low value of
$Z/X$. Solar models constructed with the GS98 composition have shown a
remarkable agreement with the solar structure, as determined by
helioseismology techniques (Christensen-Dalsgaard et al. 1996; Bahcall
et al. 1997; Morel et al. 1999; Basu et al. 2000). However, when the
AGS05 composition is adopted in the solar models, the predicted
surface helium abundance is too low and the convective envelope too
shallow.  Additionally, the model sound-speed and density profiles
show a degraded agreement with their solar counterparts when compared
to predictions from models that use the older GS98 composition
(Bahcall \& Pinsonneault 2004; Basu \& Antia 2004; Bahcall et
al. 2005a, 2005b, 2005c, 2006; Delahaye \& Pinsonneault 2006).

The discrepancy between the low-$Z/X$ models and the helioseismic
observations has led to attempts to determine the solar metallicity
from helioseismic data alone, just as the solar helium abundance was
determined by helioseismology. Antia \& Basu (2006) used helioseismic
data to estimate a value for $Z$ in the solar convection zone of
$0.0172\pm 0.002$, i.e., closer to the GS98 value and much larger than
the AGS05 value. The uncertainty in their results arose from
uncertainties in the equation of state, and a lack of data on acoustic
modes of high angular degree ($ l > 200$).  In this paper we look to
the solar core, where uncertainties in the physics of the equation of
state and opacities are much lower, to try to constrain the solar
metallicity.

We make use of solar $p$-mode data derived from observations made by
the ground-based Birmingham Solar-Oscillations Network (BiSON; Chaplin
et al. 1996). The BiSON instruments make disc-averaged observations of
the Sun in Doppler velocity.  BiSON data can be used to determine very
precise frequencies of low-$l$ modes ($l \le 3$) that can be used to
probe the solar core. The possibility of using these data to shed
light on the solar abundance problem was explored by Basu et
al. (2007). Basu et al. made very specific combinations of the low-$l$
frequencies, the so-called `small frequency spacings' and `frequency
separation ratios', to compare models with the observations.

The \emph{small frequency spacings} of the low-$l$ $p$ modes are given
by the combination.
 \begin{equation}
 d_{l\,l+2}(n) = \nu_{n,l} - \nu_{n-1,l+2},
 \label{eq:fine}
 \end{equation}
where $\nu_{n,l}$ is the frequency of a mode of degree $l$ and radial
order $n$.  The fine spacings are determined predominantly by the
sound-speed gradient in the core. Using the asymptotic theory of $p$
modes it can be shown that (see e.g., Christensen-Dalsgaard \&
Berthomieu 1991)
\begin{equation}
d_{l\,l+2}(n)\simeq -(4l+6){\Delta_l(n)\over{4\pi^2\nu_{n,l}}}\int_0^R
{dc\over dr}{dr\over r},
\label{eq:smallsep}
\end{equation} 
where $R$ is the solar radius, and $\Delta_l(n)$ is the large
frequency spacing given by
 \begin{equation}
 \Delta_l(n) = \nu_{n,l} - \nu_{n-1,l}.
\label{eq:largesep}
 \end{equation}
The large frequency separation depends inversely on the sound-travel
time between the center and the surface of the Sun.  The frequencies
$\nu_{n,l}$ and $\nu_{n-1,l+2}$ are very similar and hence are
affected in a similar way by near-surface effects.  By taking this
difference in frequency a large part of the effects from the
near-surface uncertainties cancels out, making the spacings a useful
probe of the deep solar interior and core. Some residual effects do
nevertheless remain.  One way of reducing the effects of the
near-surface errors is to use the \emph{frequency separation ratios}.
The frequency separation ratios (Roxburgh \& Vorontsov 2003; Ot\'i
Floranes, Christensen-Dalsgaard \& Thompson 2005; Roxburgh 2005) are
formed from the small frequency spacings and large frequency spacings
of the modes.  The separation ratios are then constructed according
to:
 \begin{equation}
 r_{02}(n) = \frac{d_{02}(n)}{\Delta_1(n)},
 ~~~~~~~~r_{13}(n) = \frac{d_{13}(n)}{\Delta_0(n+1)}.
 \label{eq:rats}
 \end{equation}
Since both the small and large spacings are affected in a similar
manner by near-surface effects, these ratios are somewhat independent
of the structure of the surface.

Basu et al. (2007) showed that small spacings and separation ratios
for models constructed with the old GS98 composition match the
observed BiSON spacings and ratios much more closely than do the
spacings and ratios of models with the lower AGS05 composition. In
short, models constructed with higher metallicities compare better
with the BiSON data than do models constructed with lower
metallicities, although the level of agreement deteriorates when the
metallicity becomes very large. This indicates that we should be able
to determine solar metallicity using the spacing and ratio data.

In this paper, we therefore expand on the work of Basu et al. (2007),
and use the small spacings and separation ratios from BiSON data to
determine the metallicity of the Sun. We compare the observed spacings
and ratios with spacings and ratios of some 12,000 solar models. The
models, which were made for an extensive Monte Carlo simulation
(Bahcall et al. 2006), account for all the relevant uncertainties
entering standard solar model calculations.  From this comparison, we
show that it is possible to place extremely tight constraints on
$\mu_{\rm c}$, the mean molecular weight averaged over the inner
20\,\% by radius (i.e., over most of the solar core) of the Sun.
Since the mean molecular weight in the core is related to the
metallicity at the surface (i.e., the convection zone), we can also
place reasonably precise constraints on the heavy element abundance,
$Z$.  Both $\mu_{\rm c}$ and $Z$ are measures of solar metallicity,
albeit for different regions of the Sun.

The rest of the paper is organized as follows. The observed data and
the models used are described in \S~\ref{sec:data}. In
\S~\ref{sec:lin}, we compare the BiSON separation ratios with the
separation ratios of two sequences of models made to test the response
of the spacings and ratios to $\mu_{\rm c}$ and $Z$.  In this section
we also show how we can obtain a seismic estimate of the solar
$\mu_{\rm c}$ and $Z$ by comparison of the BiSON and model spacings
and ratios.  In \S~\ref{sec:op} we determine the systematic errors
that arise due to uncertainties in the radiative opacities and the
equation of state --- two quantities are that not amenable to a Monte
Carlo type study.  In \S~\ref{sec:mc} we expand the BiSON-model
comparison by analyzing the results for a grand total of 12,000 Monte
Carlo solar models, computed by Bahcall et al. (2006). This analysis
allows us to test the impact of changes to several input solar model
parameters (in particular changes to the mixture of the heavy
elements) on the results.  Finally, we summarize our results in
\S~\ref{sec:summ}.

\section{Data and models used}
\label{sec:data}

We have made use of Doppler velocity observations made by the BiSON
over the 4752-d period beginning 1992 December 31, and ending 2006
January 3. Frequencies were determined by fitting resonant peaks in
the power spectrum of the complete time series to yield estimates of
the low-$l$ frequencies (e.g., see Chaplin et al. 1999). Prior to
calculation of the small spacings and separation ratios, we removed
the solar-cycle shifts from the raw fitted low-$l$
frequencies. Details on the process used to remove the solar-cycle
shifts can be found in Basu et al. (2007) (see also Chaplin et
al. 2005). The BiSON spacings and ratios were then constructed from
these corrected frequency data, and the uncertainties on individual
fitted frequencies propagated in the usual manner to give the spacing
and ratio uncertainties. Table~\ref{tab:bison} contains those
solar-cycle-corrected BiSON mode frequencies which were used to
compute the BiSON frequency spacings and separation ratios analyzed in
the paper.

Basu et al. (2007) showed that the separation ratios depend on the
molecular weight, but they did not determine the exact dependence of
the separation ratios on the average mean molecular weight of the
core. To do so, we use two very different sets of models. The first
set of models, which we refer to as the \emph{test} models, consist of
two sequences of ten solar models each. One sequence of models was
constructed with the relative heavy element abundances of GS98, while
the second sequence was made with the relative heavy abundances of
AGS05.  To fix the $Z/X$ of a given model in either sequence, the
individual relative heavy element abundances of GS98 (or AGS05) were
multiplied by the same constant factor. This factor was then changed
from one model to another within the sequence.  All models in the two
sequences were constructed with the same nuclear reaction rates,
opacities, equation of state and diffusion rates.  These models were
constructed to test the dependence of the separation ratios on the
\emph{average} mean molecular weight, $\mu_{\rm c}$ and the
\emph{total} heavy element abundance, $Z$. For reference, we include
in this paper tables of low-$l$ mode frequencies for two standard
models. Table~\ref{tab:gs98} has frequencies for a model with the
exact GS98 abundance; while Table~\ref{tab:ags05} has frequencies for
a model with the exact AGS05 abundance.

The second set of models comprised a grand total of 12,000 models
created for a Monte Carlo study (Bahcall et al. 2006), and we refer to
these as the \emph{Monte Carlo} models.  The characteristics and
methods of computation of the models can be found in Bahcall et
al. (2006). Here, we summarize the salient points only.

For each solar model 19 input parameters were drawn randomly from
separate probability distributions for every parameter (see Bahcall et
al. 2006 for more details).  Seven of the input parameters were
nuclear reaction rates for important low-energy fusion reactions. The
solar age, luminosity, and diffusion coefficient (rate) were the next
three parameters. The final nine parameters were the abundances of
nine heavy elements: C, N, O, Ne, Mg, Si, S, Ar and Fe. For each solar
model, radiative opacity tables corresponding to the randomly
generated composition were computed and used. Radiative opacity tables
were generated using data and codes provided by the Opacity Project
group as presented in Seaton (2005). Low-temperature opacities (for
temperatures under $10^4\,\rm K$) were from Ferguson et al. (2005).
Variations of the abundances were dealt with in such a way as to give
four sequences of Monte Carlo models. Choices had to be made regarding
the underlying mixture, and the probability distribution for the
mixture. Two basic mixtures were used: The GS98 and the AGS05
mixtures. Probability distributions were then assigned on the basis of
two different estimates of the uncertainties in the abundances of the
nine individual elements: `conservative' (large) uncertainties, based
on differences between the abundances of the GS98 and AGS05 mixtures;
and `optimistic' (small) uncertainties, based on the uncertainties
quoted by Asplund et al.  (2005a). The content of the four sequences
of models may be summarized as follows:

GS-Cons --- These 5000 models were made with the `conservative'
(large) abundance uncertainties, centered on the GS98 mixture.

GS-Opt --- These 1000 models were made with the `optimistic'(small)
abundance uncertainties, centered on the GS98 mixture.

AGS-Cons --- These 1000 models were made with the `conservative'
(large) abundance uncertainties, centered on the AGS05 mixture.

AGS-Opt --- These 5000 models were made with the`optimistic' (small)
abundance uncertainties, centered on the AGS05 mixture.

We have calculated frequencies of low-degree modes for all the models
and constructed the small spacings and separation ratios in exactly
the same manner as for the observations. The model frequencies come
from adiabatic calculations. Since non-adiabatic effects are not
included, this leads to a well-known mismatch between the absolute
values of the adiabatic model frequencies and the observed
frequencies. However, as noted earlier, differences due to these
near-surface effects are reduced significantly by taking frequency
differences, and using the small spacings and separation ratios.

\section{Dependence of the separations on metallicity}
 \label{sec:lin}

In order to parametrize the relation between metallicity and the small
spacings and separation ratios, we begin by comparing the BiSON
spacings and ratios with the spacings and ratios of the \emph{test}
models. To show how the BiSON-model comparisons were made, consider
the analysis of the separation ratios. We calculated for each model
the differences between the observed BiSON ratios, $r_{02}(n)$ and
$r_{13}(n)$, and the model ratios, $r'_{02}(n)$ and $r'_{13}(n)$,
i.e.,
 \begin{equation}
 \Delta r_{l,l+2}(n) = r_{l,l+2}(n) - r'_{l,l+2}(n).
 \label{eq:diff1}
 \end{equation}
These differences were then averaged over $n$, for each of the $\Delta
r_{02}(n)$ and $\Delta r_{13}(n)$, to yield \emph{weighted mean
differences}, $\langle \Delta r_{02} \rangle$ and $\langle \Delta
r_{13} \rangle$:
 \begin{equation}
 \langle \Delta r_{l,l+2} \rangle =
 \frac{\displaystyle\sum_{n}^{} \Delta r_{l,l+2}(n) / \sigma^2_{r_{l,l+2}}(n)}{
       \displaystyle\sum_{n}^{} 1/\sigma^2_{r_{l,l+2}}(n)}.
 \label{eq:diff2}
 \end{equation}
The formal uncertainties of the BiSON spacings,
$\sigma_{r_{l,l+2}}(n)$, were used to weight the averages (with the
usual uncertainty-squared Gaussian weighting applied). We averaged
data over the ranges where good determinations of the separation
ratios were available, here $n=9$ to 25.

\begin{figure*}
 \epsscale{1.0}
 \plottwo{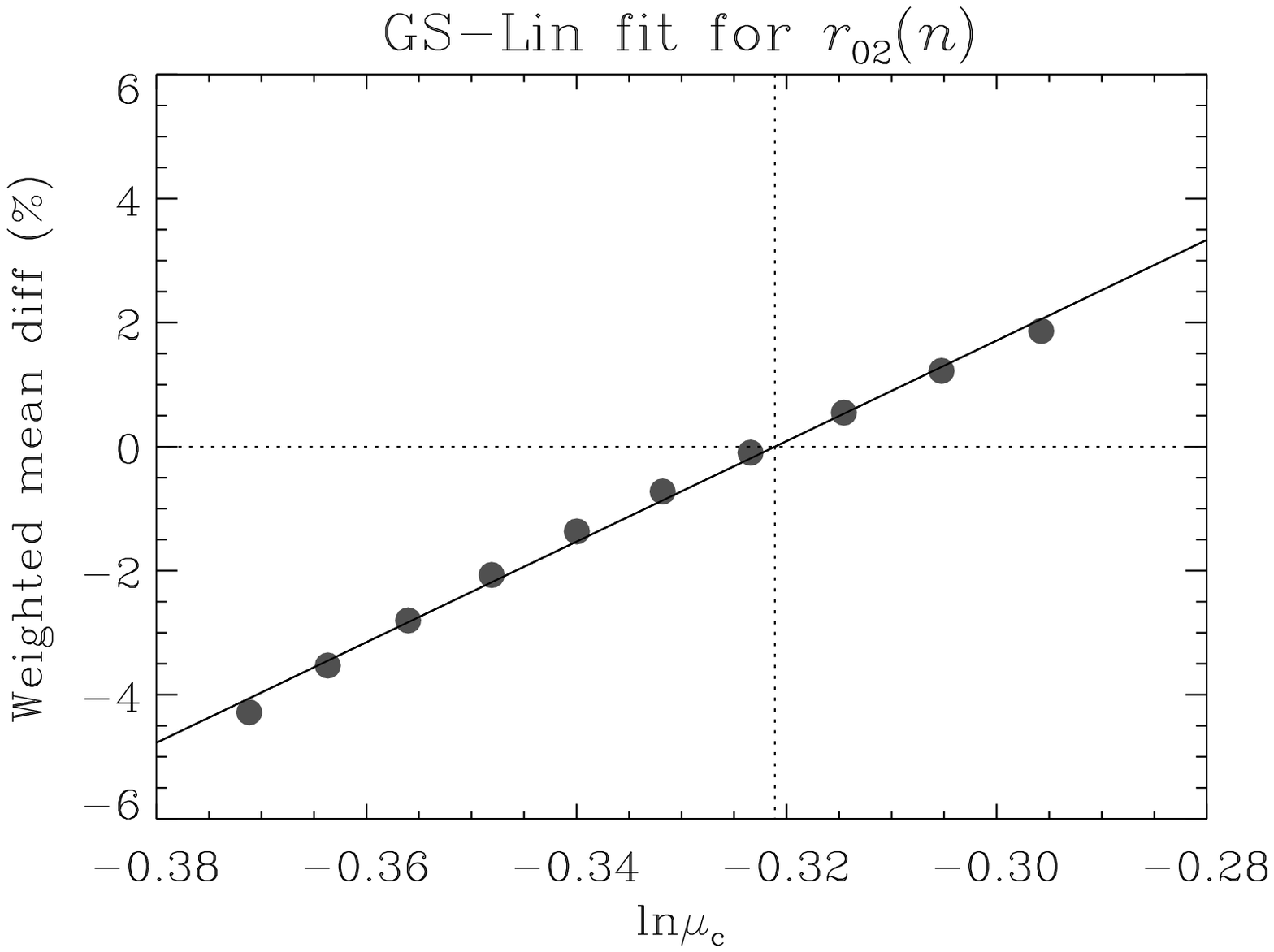}{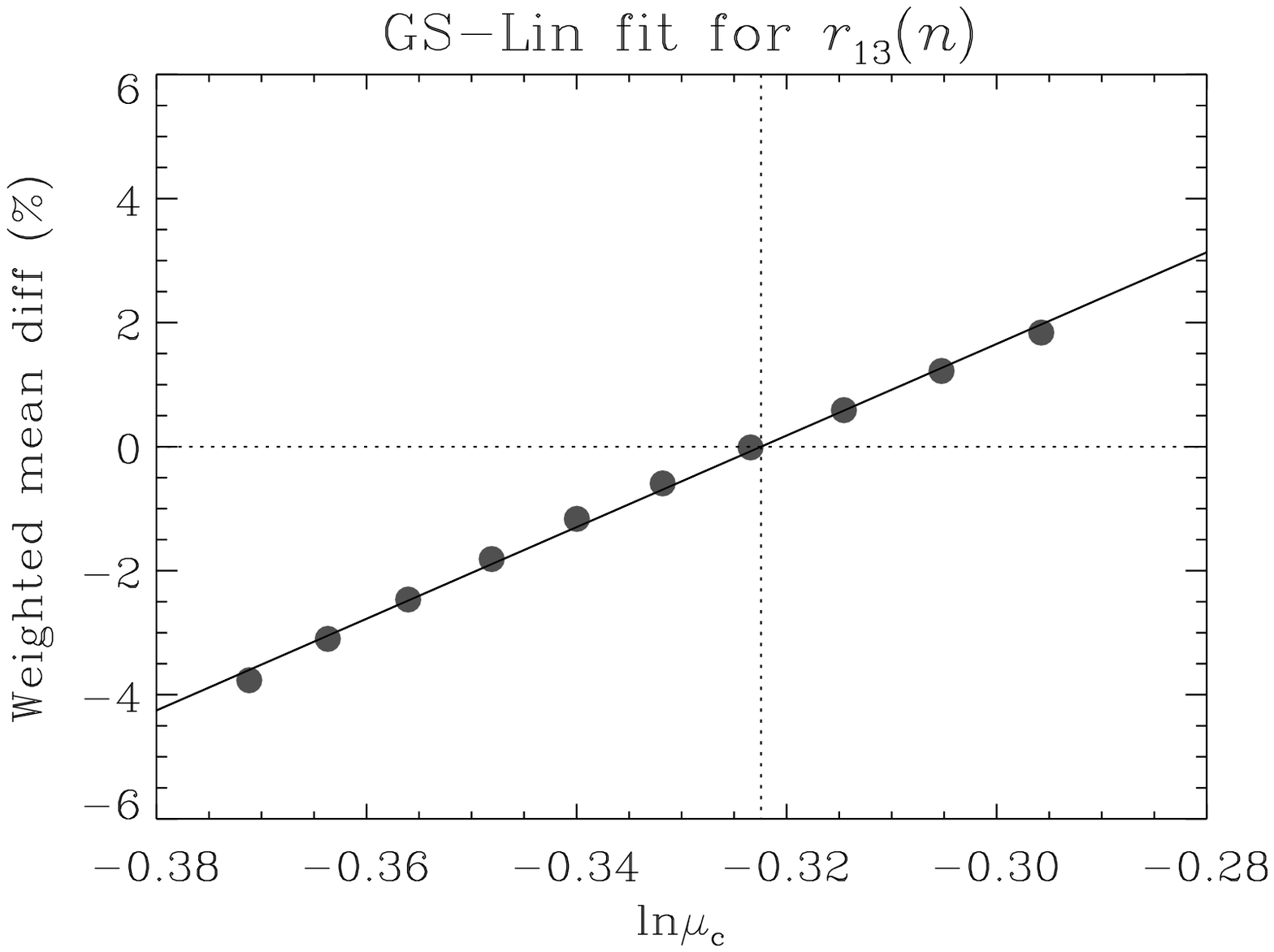}\\
 \plottwo{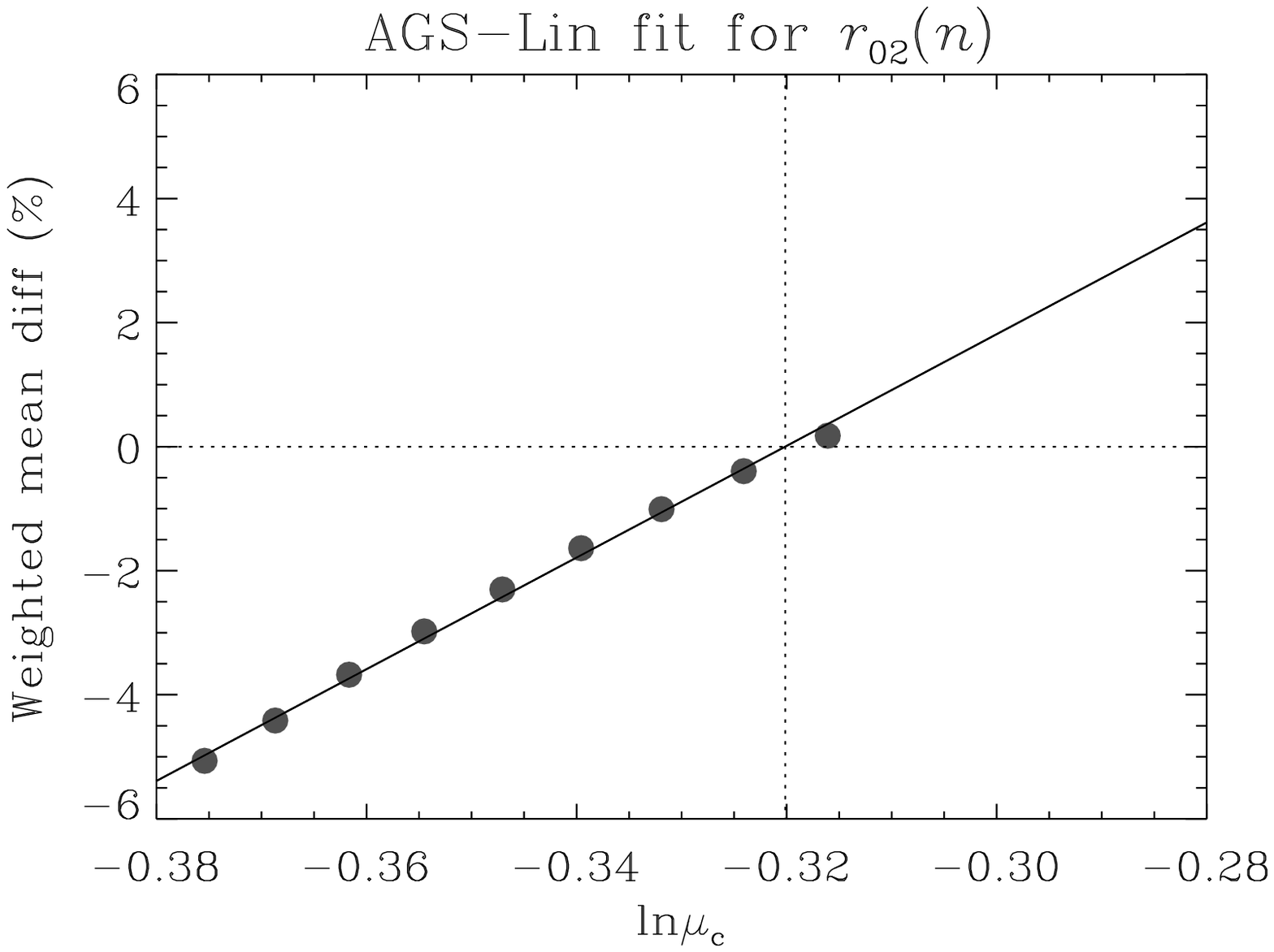}{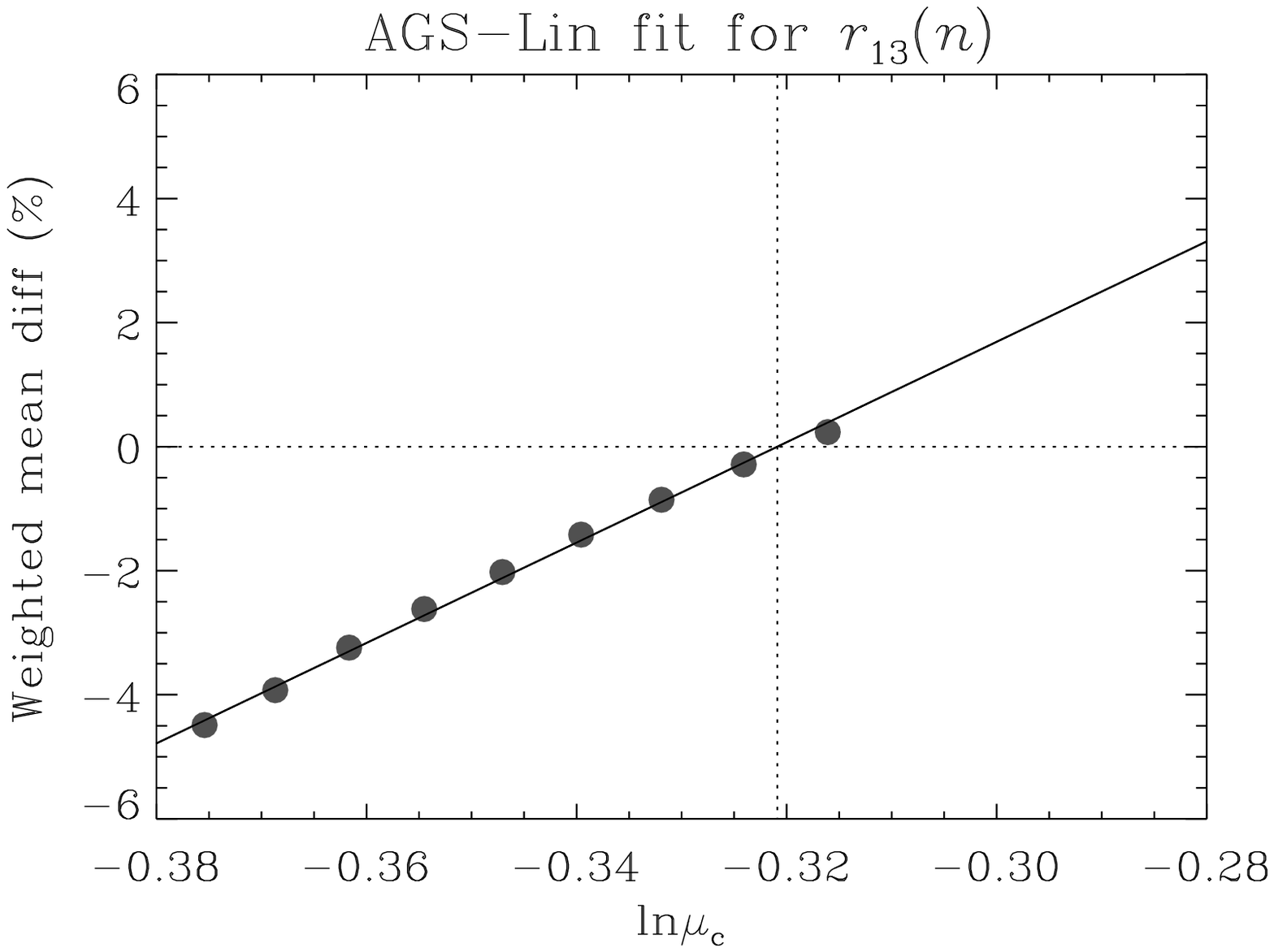}

\caption{Upper two panels: weighted mean differences, $\langle \Delta
r_{02} \rangle$ (upper left-hand panel) and $\langle \Delta r_{13}
\rangle$ (upper right-hand panel), versus $\ln\mu_{\rm c}$, plotted
for a sequence of 10 solar models. All models have the same relative
mixture of heavy elements, corresponding the Grevesse \& Sauval (1998)
mixture but differ in the total amount of metals.  The solid line in
each panel is the best-fitting straight line. The dotted lines
intersect at the location along each best-fitting line where the
weighted mean difference is zero. Lower panels: as per upper panels,
but for a sequence of 10 solar models having a relative mixture of
heavy elements corresponding to the Asplund et al. (2005) mixture.}

 \label{fig:linmu}
 \end{figure*}

\begin{figure*}
 \epsscale{1.0}
 \plottwo{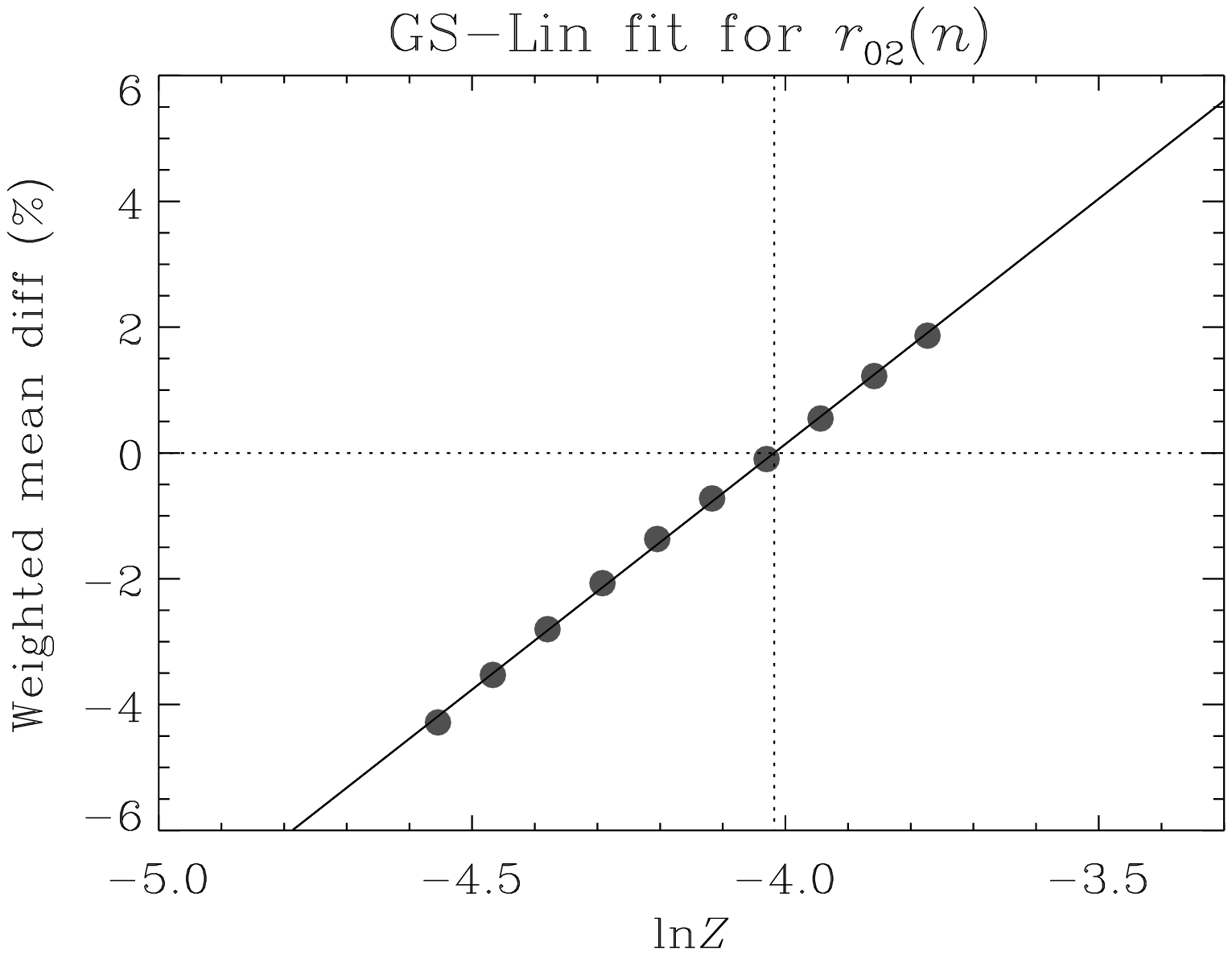}{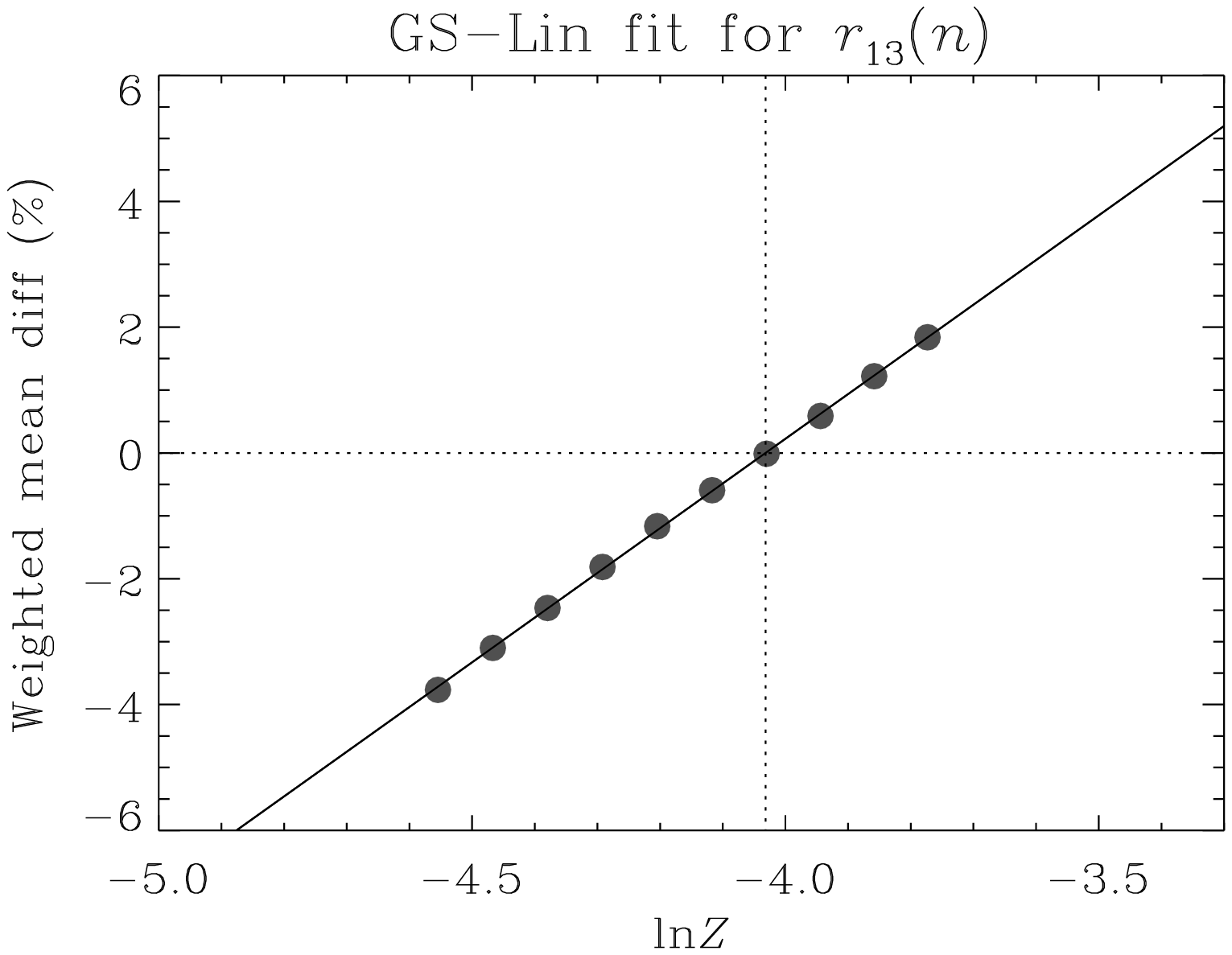}\\
 \plottwo{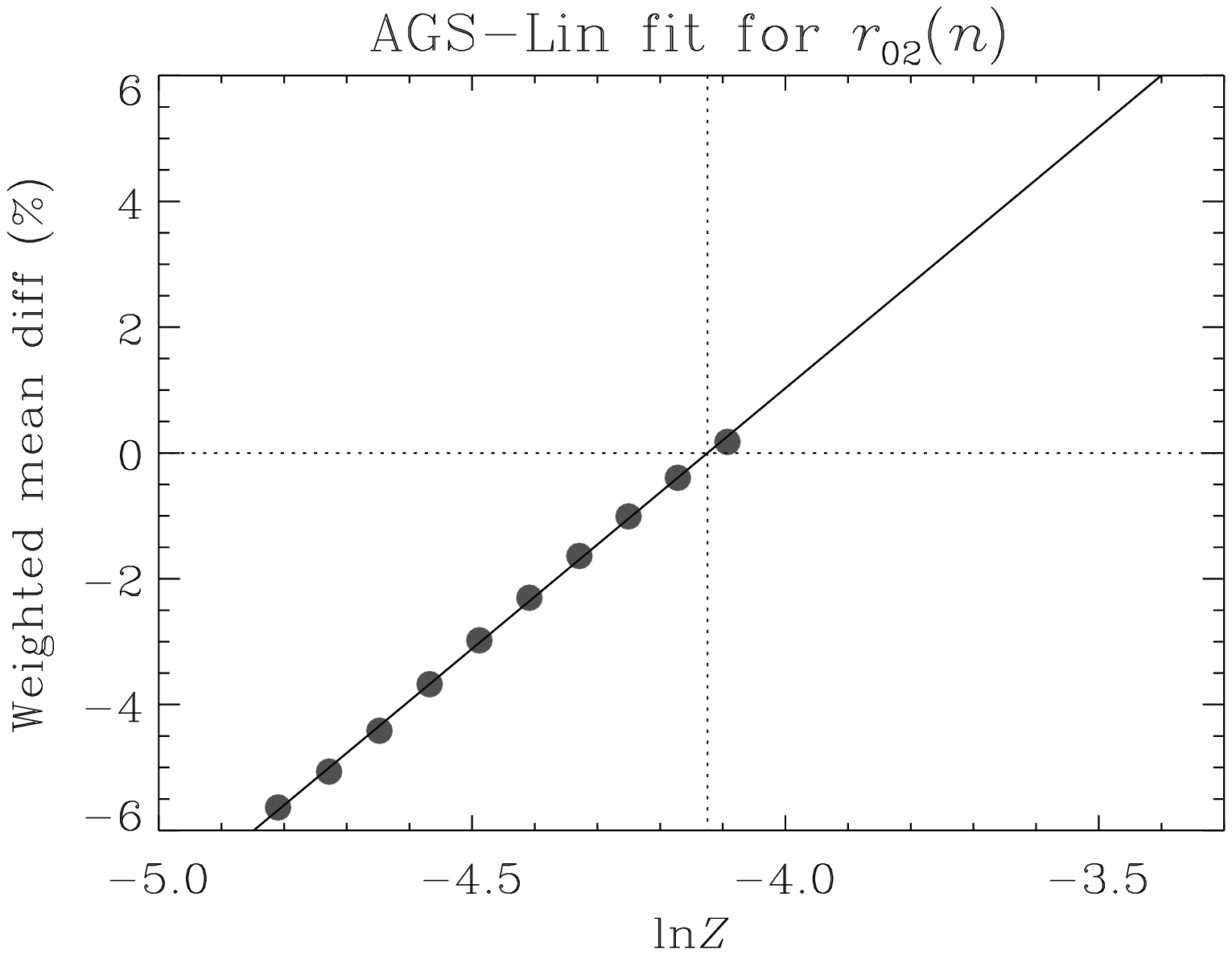}{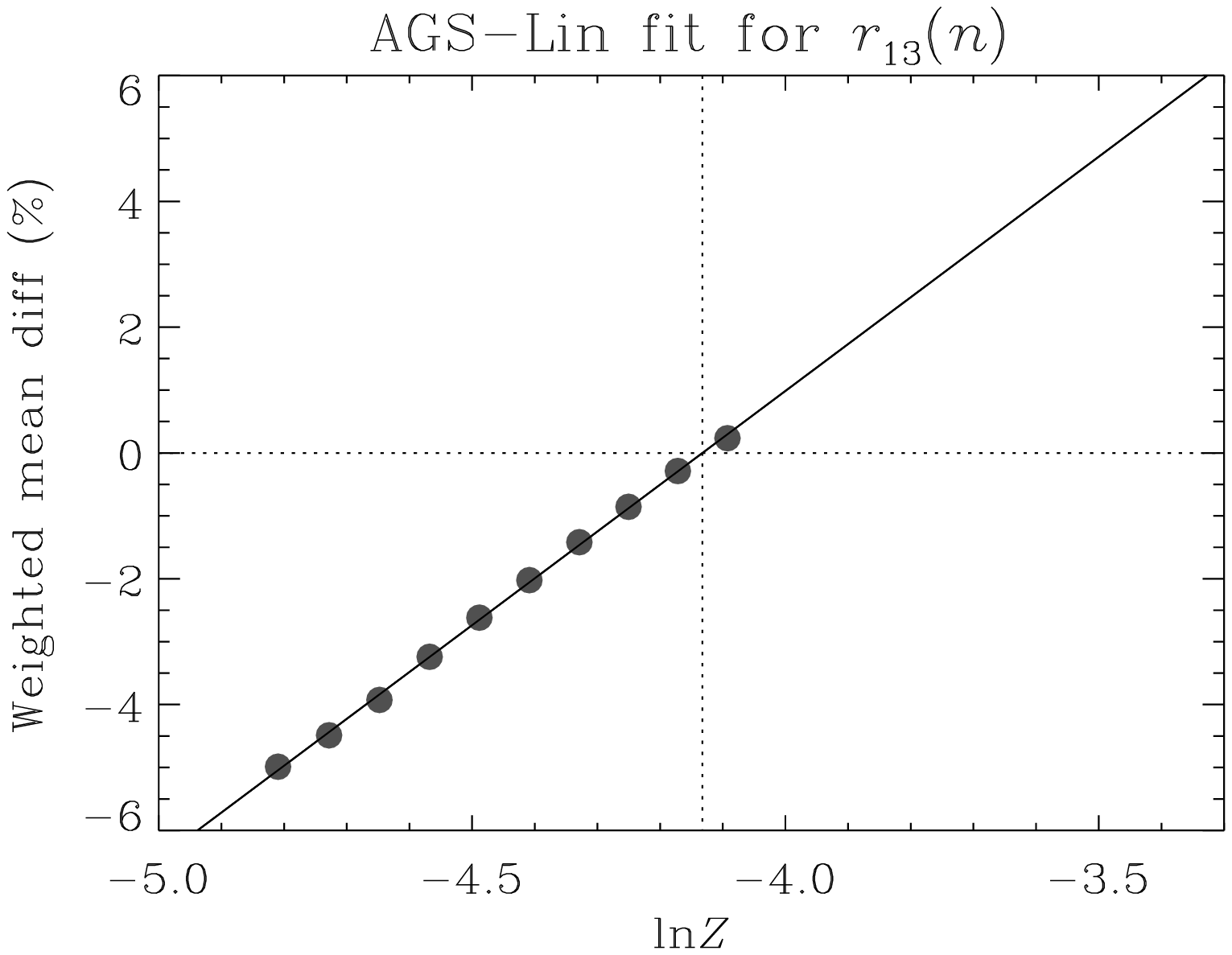}

 \caption{Upper two panels: weighted mean differences, $\langle \Delta
   r_{02} \rangle$ (upper left-hand panel) and $\langle \Delta r_{13}
   \rangle$ (upper right-hand panel), versus $\ln Z$, plotted for a
   sequence of 10 solar models. All models have the same relative
   mixture of heavy elements, corresponding the Grevesse \& Sauval
   (1998) mixture but differ in the total amount of metals. The solid
   line in each panel is the best-fitting straight line. The dotted
   lines intersect at the location along each best-fitting line where
   the weighted mean difference is zero. Lower panels: as per upper
   panels, but for a sequence of 10 solar models having a relative
   mixture of heavy elements corresponding to the Asplund et
   al. (2005) mixture.}

 \label{fig:lin}
 \end{figure*}

Fig.~\ref{fig:linmu} shows plots of the weighted mean differences of
the separation ratios (in \%) versus $\ln \mu_{\rm c}$, the natural
logarithm of the average mean molecular weight of the core.  The top
two panels show data for the GS98 set, and the bottom two panels show
data for the AGS05 set. The formal uncertainties on each point, which
come from the BiSON data, are not plotted. They are $0.046\,\%$ on
each $\langle \Delta r_{02} \rangle$; and $0.038\,\%$ on each $\langle
\Delta r_{13} \rangle$. Fig.~\ref{fig:lin} shows plots of the weighted
mean differences versus the natural logarithm of the surface heavy
element abundances, $\ln Z$. Again, the top two panels show data for
the GS98 set, and the bottom two panels show data for the AGS05 set.

When a straight line was fitted to the data in each plot (solid
lines), the fitting coefficients indicated that a linear dependence
was a good model for the data. The fit for the $Z$ data was in all
cases described by:
  \begin{equation}
  \langle \Delta r_{l,l+2} \rangle =
  \alpha_{l,l+2} + \beta_{l,l+2} \ln Z,
  \label{eq:fit}
  \end{equation}
where $\alpha_{l,l+2}$ and $\beta_{l,l+2}$ are, respectively, the
best-fitting estimates of the intercept and gradient of the straight
line.  A similar straight-line fitting model was used for the
$\mu_{\rm c}$ data, with $\ln \mu_{\rm c}$ used as the independent
variable.

The quality of the $Z$ fits degraded significantly when $Z$, rather
than $\ln Z$, was used as the independent variable. Visible departures
from a straight line were then observed.  The fits also deteriorated
when the fine spacings, rather than the separation ratios, were used.
We therefore devote the remainder of the paper to analysis of weighted
mean differences made from the separation ratios of the BiSON data and
the solar models.  These weighted mean differences have then been used
to determine the natural logarithm of $Z$, as opposed to $Z$
itself. We adopted a similar approach to our study of the mean
molecular weight in the core. Here, we used the $\ln \mu_{\rm c}$ of
the models as the independent variable for the plots. Again, this was
because we found that use of $\mu_{\rm c}$, rather than $\ln \mu_{\rm
c}$, degraded the quality of the fits somewhat, although not as much
as in the case of $Z$.

With reference to Figures~\ref{fig:linmu} and~\ref{fig:lin}, it is not
surprising that the two measures of metallicity affect the separation
ratios in similar ways.  The quantities $\mu_{\rm c}$ and $Z$ are
related, in that a higher $Z$ results in a higher $\mu_{\rm c}$. Two
models with the same $Z$, nuclear reaction rates, opacities and
equation of state can have different values of $\mu_{\rm c}$ only if
the diffusion rates are different in the two models.  It is, however,
not surprising that the dependence of the separation ratios on the two
parameters is somewhat different given that $\mu_{\rm c}$ also depends
heavily on the helium abundance in the core. All the models are
calibrated to have the same radius and luminosity at the solar age,
and hence differences in $Z$ generally give rise to differences in the
core helium abundance.

 \subsection{Seismic estimates of solar $Z$ and $\mu_{\rm c}$}
 \label{subsec:seis}

If the observational data are unbiased, the location on the abscissa
that marks where each best-fitting straight line passes through zero
on the ordinate will give us a `seismic' estimate of the average mean
molecular weight $\mu_{\rm c}$ averaged over the inner 20\,\% by
radius, and the surface abundance $Z$ of the Sun.  (Note that the $Z$
we refer to is always the present-day (solar age) surface $Z$.) These
locations are marked on the various panels of Figures~\ref{fig:linmu}
and~\ref{fig:lin} by the intersecting dotted lines. Conclusions drawn
from the test models will of course neglect any dependence of the
differences of the separation ratios on changes to other solar model
input parameters, including changes to the mixture of the heavy
elements. We go on to discuss the impact of such changes, and the
overall error budget, in later sections. Here we show explicitly how a
value for, and uncertainty on, $Z$ may be estimated from the
differences. The same procedures give results for $\mu_{\rm c}$.

From the best-fitting coefficients, we seek to find $\ln Z$ where
$\langle \Delta r_{l,l+2} \rangle=0$.  From Equation~\ref{eq:fit}, we
therefore have:
  \[
  \ln Z =  -\alpha_{l,l+2} / \beta_{l,l+2},
  \]
So that
  \begin{equation}
  Z = \exp[-\alpha_{l,l+2} / \beta_{l,l+2}].
  \label{eq:getz}
  \end{equation}
Application of Equation~\ref{eq:getz} to the $\langle \Delta r_{02}
\rangle$ and $\langle \Delta r_{13} \rangle$ data of each set (GS98 or
AGS05) will give us four seismic estimates of the solar heavy element
abundance. Using GS98 models we get $Z=0.01798$ and $Z=0.01774$ for
$\langle \Delta r_{02} \rangle$ and $\langle \Delta r_{13} \rangle$
respectively. The corresponding results using the AGS05 models are
$Z=0.01617$ and $Z=0.01611$. A similar analysis of $\mu_{\rm c}$
results in values of $0.7253$ and $0.7244$ when GS98 models are used,
and $0.7260$ and $0.7255$ when AGS05 test models are used.  These
estimates are listed in the fourth column of Tables~\ref{tab:linmu}
and \ref{tab:lin}, along with estimates of the goodness-of-fit and
uncertainties.

We make use of the observed scatter (variance) of the differences
$\langle \Delta r_{02} \rangle$ and $\langle \Delta r_{13} \rangle$
about their best-fitting straight line to estimate the uncertainty on
the seismic estimates of solar $\mu_{\rm c}$ and $Z$.  In short, we
translate the characteristic scatter on the ordinate into an implied
uncertainty on the abscissa. To determine the uncertainty in $Z$ (the
same procedure was used for $\mu_{\rm c}$) we first determine the set
of residuals about the best-fitting straight line, i.e., for each
point we compute
  \[
  \delta \langle \Delta r_{l,l+2} \rangle =
 \langle \Delta r_{l,l+2} \rangle - [\alpha_{l,l+2} + \beta_{l,l+2} \ln Z].
  \]
The variance of these residuals yields an estimate of their $1\sigma$
standard deviation, which we call $\sigma_{\langle \Delta r_{l,l+2}
\rangle}$. This characteristic uncertainty on the residuals may be
translated into an implied uncertainty on $\ln Z$ via the best-fitting
gradient, i.e.,
  \begin{equation}
  \sigma(\ln Z) = \sigma_{\langle \Delta r_{l,l+2} \rangle} /
  \beta_{l,l+2}.
  \label{eq:method2}
  \end{equation}
The equivalent $1\sigma$ limits on $Z$ are then bounded by $\exp{[\ln
Z+\sigma(\ln Z)]}$ and\\ $\exp{[\ln Z-\sigma(\ln Z)]}$. Our estimate
of the uncertainty is itself uncertain through the uncertainty on the
gradient, $\beta_{l,l+2}$. Thus we require that $N$ be large enough to
ensure that the best-fitting gradient (and our `look-up curve') is
well constrained. Here, gradients for the $Z$ fits were returned to a
fractional precision of better than 1\,\%.

The third column of Tables~\ref{tab:linmu} and~\ref{tab:lin} shows the
computed $\sigma_{\langle \Delta r_{l,l+2} \rangle}$ (in \%). The
fifth column gives the implied $1\sigma$ uncertainties on $\mu_{\rm
c}$ and $Z$, which we call $\sigma(\mu_{\rm c})$ and $\sigma(Z)$.
Here, the positive and negative uncertainties were the same, at the
level of precision of the data, because the fractional uncertainties
were so small. The sixth column shows the implied precision in the
determination of $\mu_{\rm c}$ and $Z$ (in \%). For each of the GS98
and AGS05 sets, we also combined the estimates from $\langle \Delta
r_{02} \rangle$ and $\langle \Delta r_{13} \rangle$ -- on the
assumption the estimates are independent -- to give the estimates
shown in the third and sixth rows of the tables.

Inspection of the $\mu_{\rm c}$ results in Table~\ref{tab:linmu} shows
that all results (both individual and combined) are consistent with
one another.  This is not surprising given the almost direct
dependence of the sound speed, and its derivative, on $\mu_{\rm c}$.  The
combined estimate obtained with the GS98 models differs from that
obtained with the AGS05 models by only $1\sigma$. As we shall see in
\S~\ref{sec:mc}, the uncertainties on the seismic estimates of
$\mu_{\rm c}$ do not increase much when other changes to the solar
model input parameters are considered.

Inspection of the $Z$ results in Table~\ref{tab:lin} shows that the
individual, and combined, estimates for solar $Z$ are all
significantly higher than the `low' Asplund et al. value of $Z \sim
0.0122$. However, our combined GS estimate ($0.01785$) and its
combined AGS counterpart ($0.01611$) differ from each other by
$\approx 16\sigma$ (combined uncertainty). This difference might at
first glance be seen as a cause for concern and an indication that
systematic errors are much larger than the random errors caused by
uncertainties in the observed frequencies. However, we go on to show
in \S~5 that other systematic sources of error -- arising from the
sensitivity of the separation ratios to other parameters of the solar
models, including the relative mixture of the heavy elements -- mean
that realistic estimates of the uncertainties on $Z$ are actually
larger in size than the uncertainties given in
Table~\ref{tab:lin}. This is in stark contrast to what is found for
$\mu_{\rm c}$ (see previous paragraph).

\begin{figure}
\plotone{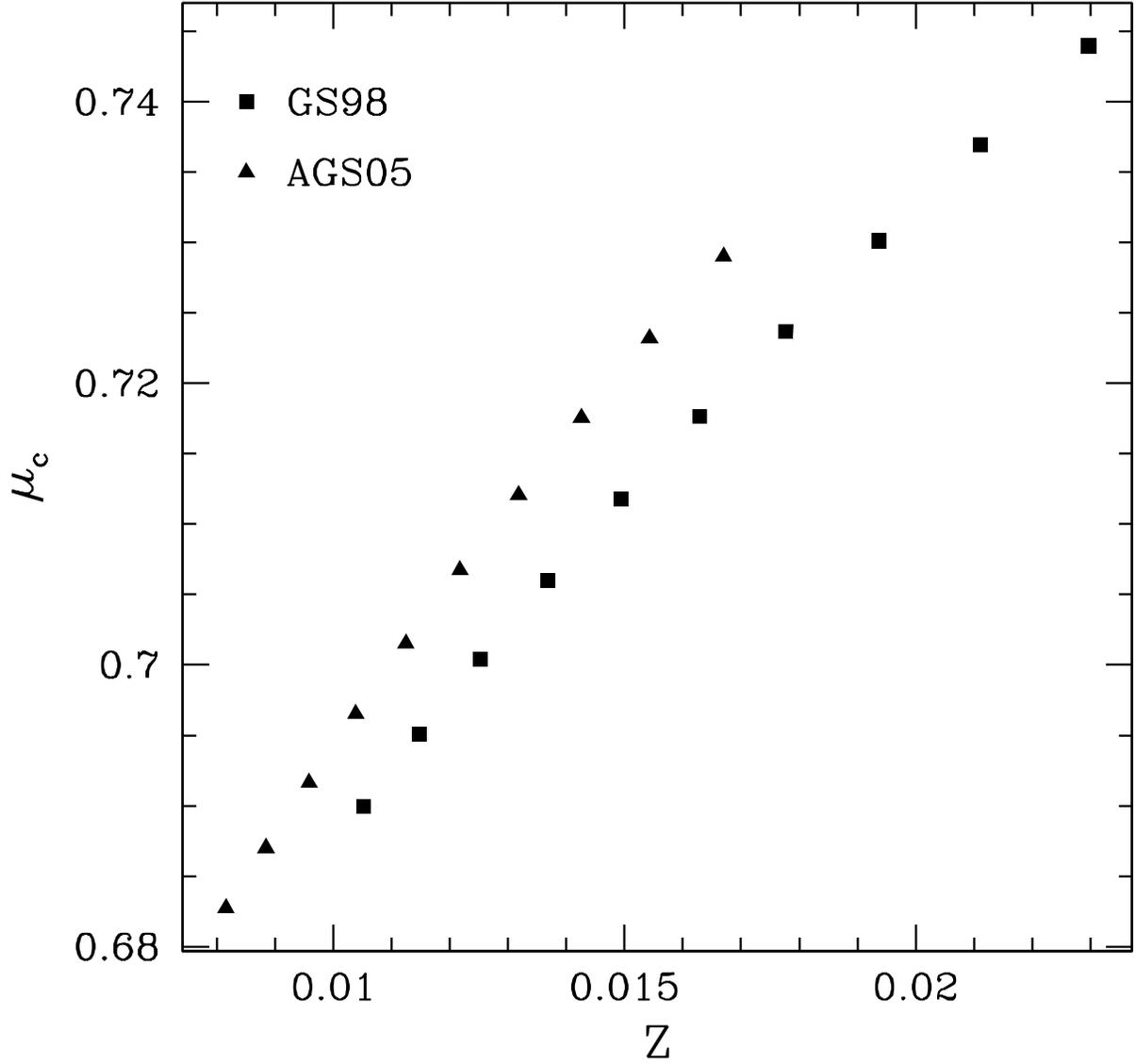}
\caption{The relation between $Z$ and $\mu_{\rm c}$ (the average mean
molecular weight in the inner 20\% by radius) for models constucted
with the GS98 relative heavy element abundances and the AGS05 relative
heavy element abundances.}
\label{fig:zmu}
\end{figure}

Disagreement between the seismic results obtained from the GS98 and
AGS98 models can to a large extent be understood in terms of the
differences in the two mixtures.  The influence of the relative
mixture is clear from the fact that the separation ratios for the
AGS05 models and the GS98 models are different for the same value of
$Z$; the differences are less pronounced for $\mu_{\rm c}$. This
finding is not difficult to understand. For the calibrated solar
models used in this work, the dominant contribution to the $Z$ and
$\mu_{\rm c}$ of each model comes from different elements. For $Z$,
the dominant elements, in order of importance, are oxygen, carbon,
neon and nitrogen.  The value of $\mu_{\rm c}$ is determined by the mass
fractions of helium and hydrogen in the core. The abundances of
hydrogen and helium in the core depend strongly on the abundances of
heavy metals that contribute to the opacity in the core. These metals,
again in order of importance, are iron, silicon, sulfur and
oxygen. The difference between the GS98 and AGS05 mixture lies
predominantly in the relative abundances of oxygen, carbon, nitrogen
and neon, and much less so in the abundances of iron, silicon and
sulfur. This explains why for the same $Z$, $\mu_{\rm c}$ is different
for the GS98 and AGS05 models, as can be seen from Fig.~\ref{fig:zmu}.

Since the separation ratios depend basically on $\mu_{\rm c}$ and
temperature, the location at which $\langle \Delta r_{l,l+2} \rangle
=0$ occurs at slightly different values for the two sets of
models. The slopes of the $\langle \Delta r_{l,l+2} \rangle$-$\ln Z$
curves for the GS98 and AGS05 models (Figures~\ref{fig:lin}) are also
different for the same reason.  We investigate other sources of
systematic errors in \S~\ref{sec:mc}. The results in \S~\ref{sec:mc}
show that realistic estimates of the uncertainties on $Z$ are actually
larger in size than the uncertainties given in Table~\ref{tab:lin}.

\section{Uncertainties due to opacity and equation of state}
\label{sec:op}

The results obtained with the test models have two obvious
limitations. First, they do not test the impact of changes to the
relative mixture of the heavy elements, except to reveal that changes
in relative abundances matter. And second, the models test the
dependence of the separation ratios on $\mu_{\rm c}$ and $Z$ for a
\emph{fixed} set of solar model input parameters, with only the $Z/X$
varied. The study above does not deal with uncertainties in the
seismic estimates of solar $\mu_{\rm c}$ and $Z$ caused by other input
parameters, such as nuclear reaction rates, diffusion rates,
uncertainties in the relative mixtures, etc.  We investigate these
effects by conducting a Monte Carlo study, which is described below in
\S~\ref{sec:mc}. This study allowed us to test the impact of 19 solar
model input parameters on the results. There are, however, two
important inputs that are not amenable to a Monte Carlo study, and
these are radiative opacities and the equation of state (EOS). These
two quantities cannot be described by a single number and hence we are
forced to use a different approach to determine the uncertainties in
the separation ratios caused by uncertainties in opacities and EOS. We
use an approach similar to that used by Bahcall et al. (2006).

We determined the uncertainty introduced in $\left< \Delta
r_{l,l+2}\right>$ by the opacities as follows. We computed a pair of
solar models with the same input parameters and EOS, but one was made
with opacities from the OP project (Badnell et al. 2005) and the other
with opacities from OPAL (Iglesias \& Rogers 1996).  For this matched
pair of models we get $\left< \Delta r_{l,l+2}\right>_{{\rm OP},i}$
and $\left< \Delta r_{l,l+2}\right>_{{\rm OPAL},i}$ (here the
subscript {\it i} denotes the pair of matched models).  The unbiased
estimator for the variance of the difference is
 \begin{equation}
 s^2_i{(\left< \Delta r_{l,l+2}\right>({\rm opacity}))}= \left[\left<
 \Delta r_{l,l+2}\right>_{{\rm OP},i} - \left< \Delta r_{l,l+2}\right>_{{\rm
 OPAL},i} \right]^2/2.
 \label{eq:var}
 \end{equation}
and we adopt this quantity as the standard deviation squared,
$\sigma^2_i{(\left< \Delta r_{l,l+2}\right>{\rm (opacity))}}$.  In
order to obtain a more representative value for $\sigma_{\rm
opac}{(\left< \Delta r_{l,l+2}\right>)}$ we averaged this difference
over a set of $N=40$ matched pairs of models, where the 19 other input
parameters were varied for different matched pairs.  The final
expression we adopt for $\sigma_{\rm opac}{(\left< \Delta
r_{l,l+2}\right>)}$ is
\begin{equation}
\sigma_{\rm opac}{(\left<   \Delta  r_{l,l+2}\right>)}=   \sqrt{N^{-1}
  \sum_i{s^2_i{(\left<  \Delta r_{l,l+2}\right>({\rm opacity}))}} }.  
\label{eq:opac}
\end{equation}
An analogous procedure was used for the EOS, but in this case one
model in each pair was computed using the 2001 OPAL EOS (Rogers 2001,
Rogers \& Nayfonov 2002), while the other model was computed using the
1996 OPAL EOS (Rogers et al.  1996).

In addition to the uncertainties on $\left< \Delta r_{02} \right>$ and
$\left< \Delta r_{13} \right>$ given by Equation~\ref{eq:opac}, we
applied the same procedure to compute the implied uncertainties for
the values of $\mu_{\rm c}$ and $Z$ predicted by the solar models.
Uncertainties in the opacity and EOS will affect the solar model
results for $Z$ and $\mu_{\rm c}$, i.e.  $Z$ and $\mu_{\rm c}$ will be somewhat
different for the two models in each of the matched pairs described
above.  The results are given in Table~\ref{tab:opac_eos}.

The impact on the separation ratios of uncertainty in the radiative
opacities is easy to understand --- changes in opacity cause changes
in temperature, which in turn change sound speed and its derivative,
thereby changing the separation ratios. However, since the opacity
uncertainties are small in the core, the overall effect is quite
modest. The impact of uncertainty in the EOS is more important. This
might seem surprising, until one realizes that the 1996 OPAL EOS did
not treat relativistic effects properly at temperatures and densities
relevant to the solar core. This results in a somewhat deficient core
structure (see Elliott \& Kosovichev, 1998). Since the deficiency is
mainly in the core, it will affect the low-$l$ separations used in
this work disproportionately and cause larger uncertainties in the
seismic estimates of solar $Z$ and $\mu_{\rm c}$.  The updated 2001
OPAL EOS has the correction put in. Thus the uncertainty in the
seismic estimates of solar $Z$ and $\mu_{\rm c}$ caused by EOS uncertainties
may be considered to be an upper limit to the EOS effects.

\section{The Monte Carlo study}
\label{sec:mc}

In this section we present results, using extended sequences of solar
models, which seek to address the influence of other input parameters
on the seismic solar $Z$ results reported in \S~\ref{subsec:seis}. The
bulk of the results come from tests on 12,000 solar models created for
a Monte Carlo study of the dependence of solar model characteristics
on different input parameters (see Bahcall et al. 2006).  As discussed
in \S~\ref{sec:data}, the models have 19 different input parameters
selected at random from a distribution of the inputs.

\begin{figure*}
 \epsscale{1.0}
 \plottwo{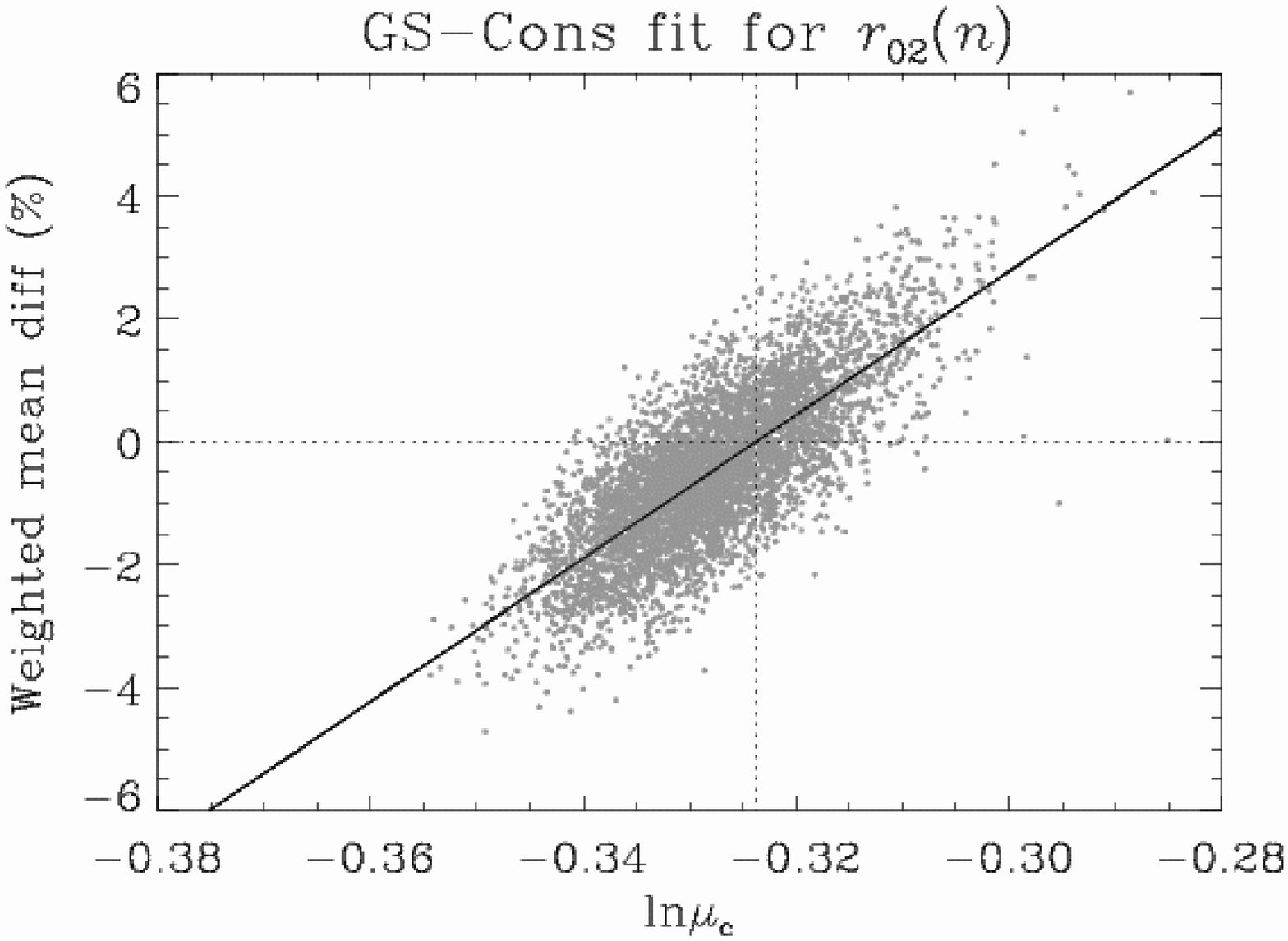}{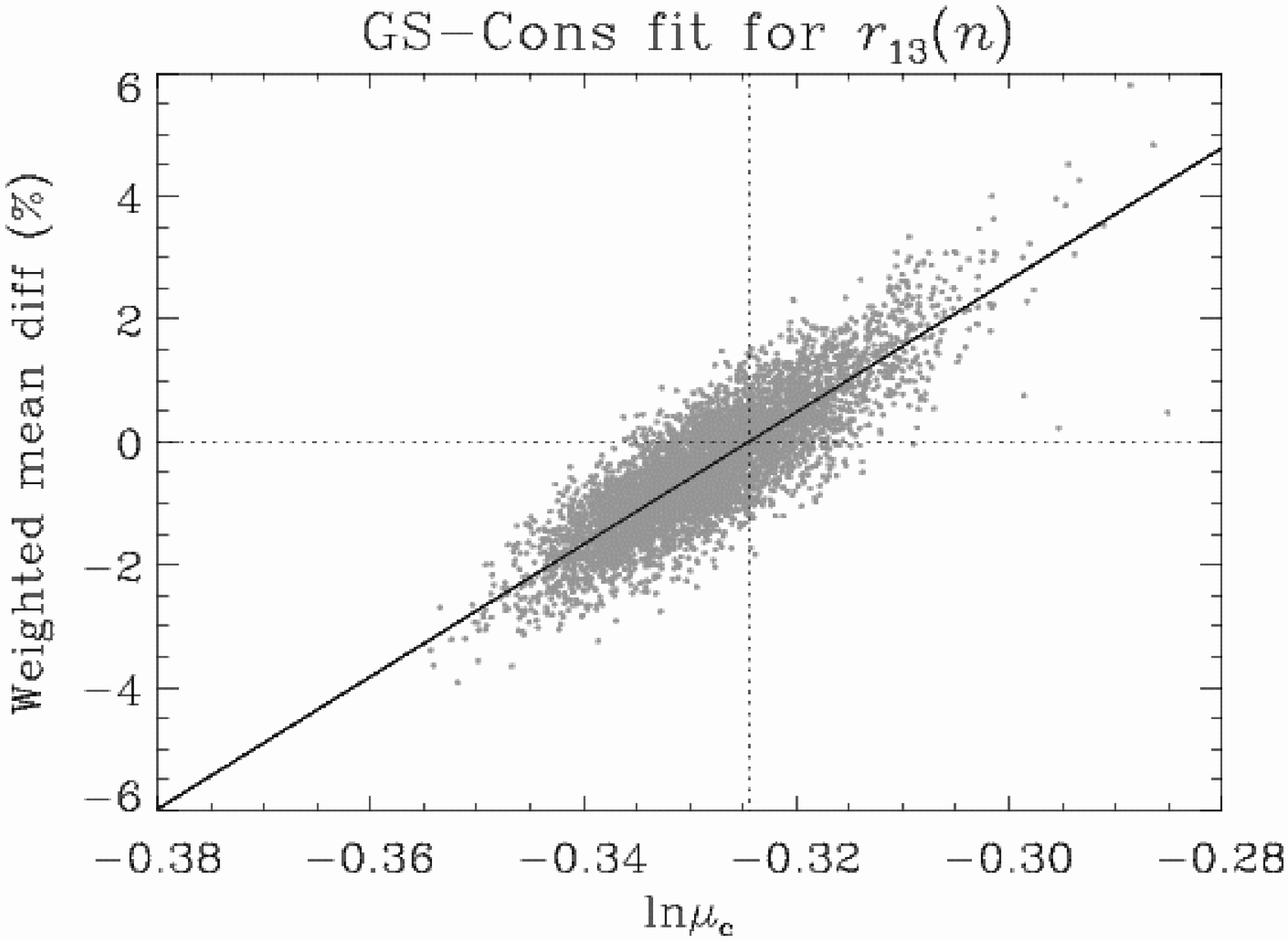}\\
 \plottwo{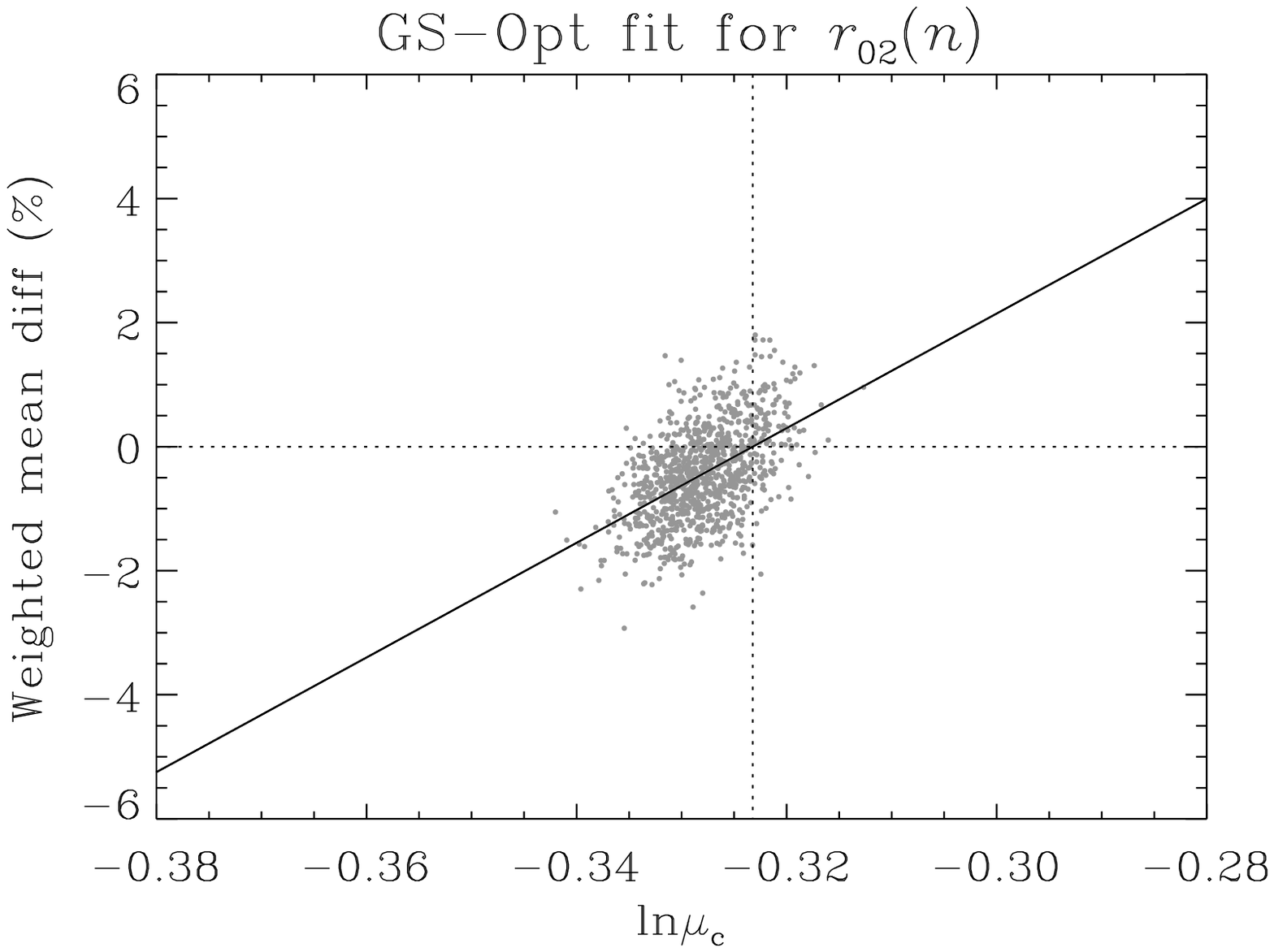}{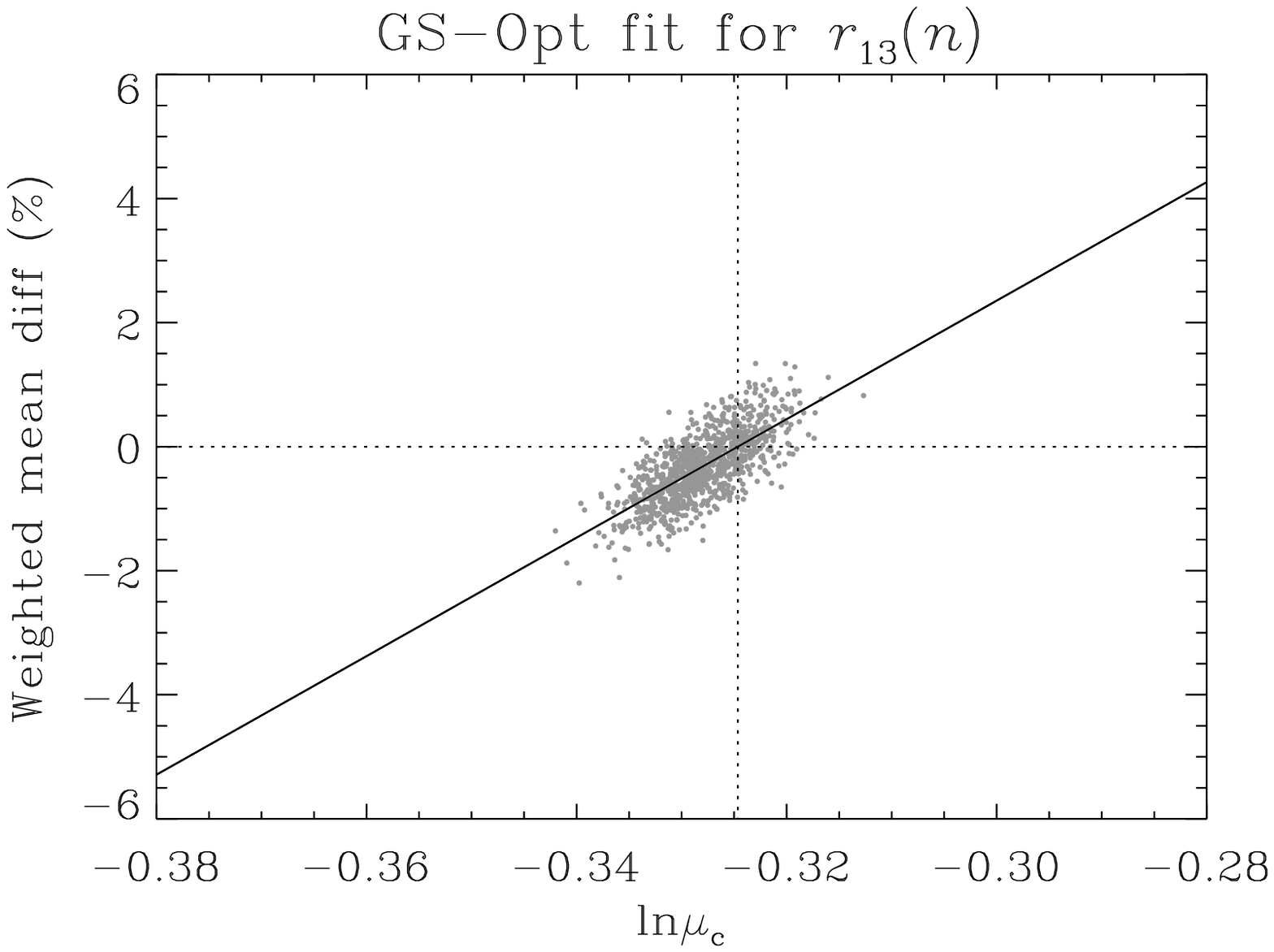}

 \caption{Upper two panels: weighted mean differences, $\langle \Delta
r_{02} \rangle$ (upper left-hand panel) and $\langle \Delta r_{13}
\rangle$ (upper right-hand panel), versus $\ln\mu_{\rm c}$, plotted
for the 5000 Monte Carlo `GS-Cons' solar models. These models were
made with the `conservative' (large) abundance uncertainties, centered
on the GS98 mixture. The solid line in each panel is the best-fitting
straight line. The dotted lines intersect at the location along each
best-fitting line where the weighted mean difference is zero. Lower
panels: As per the upper panels, but for the 1000 Monte Carlo `GS-Opt'
solar models. These models were made with the `optimistic' (small)
abundance uncertainties, centered on the GS98 mixture.}

 \label{fig:GSmuMC}
 \end{figure*}

\begin{figure*}
 \epsscale{1.0}
 \plottwo{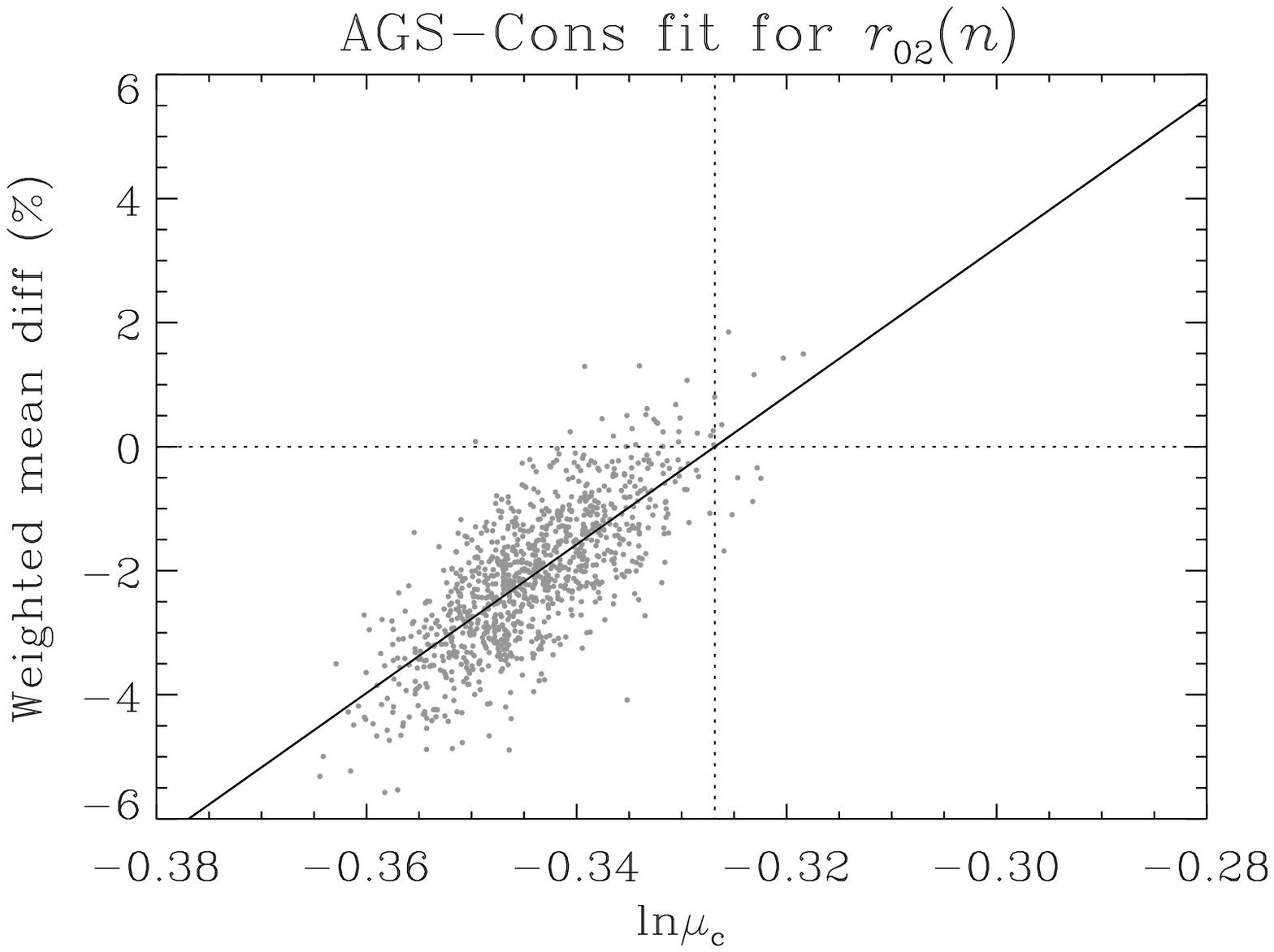}{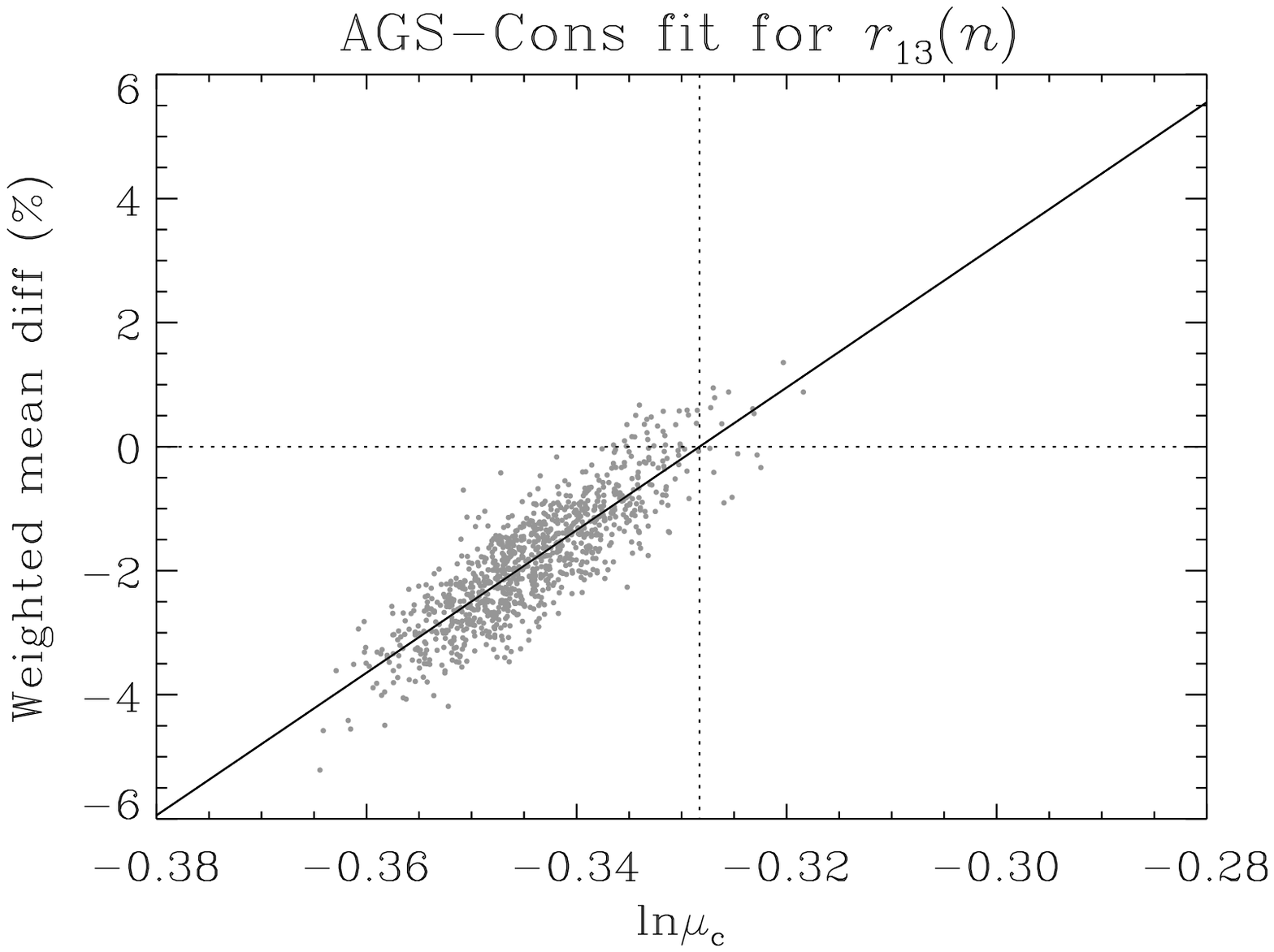}\\
 \plottwo{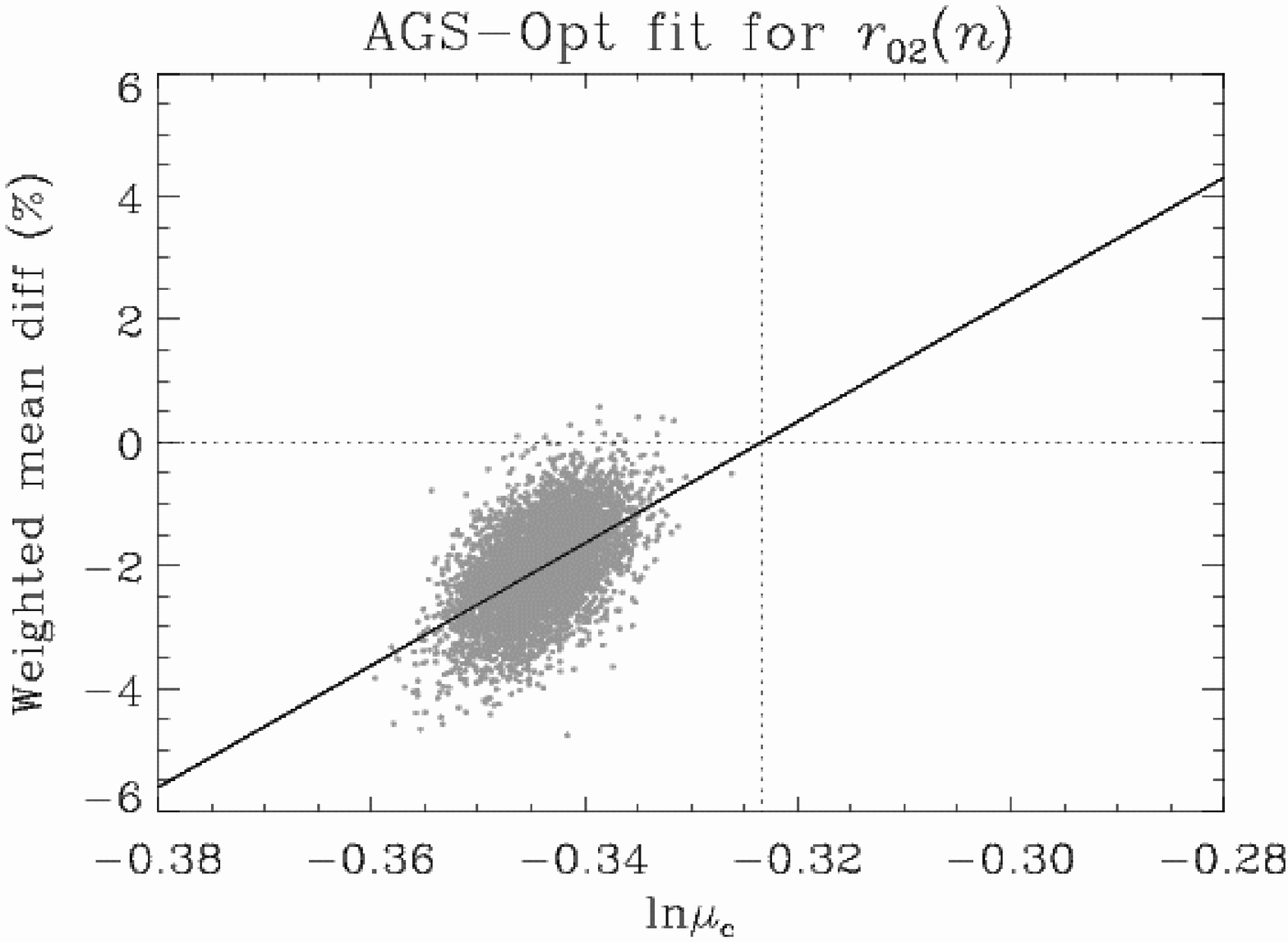}{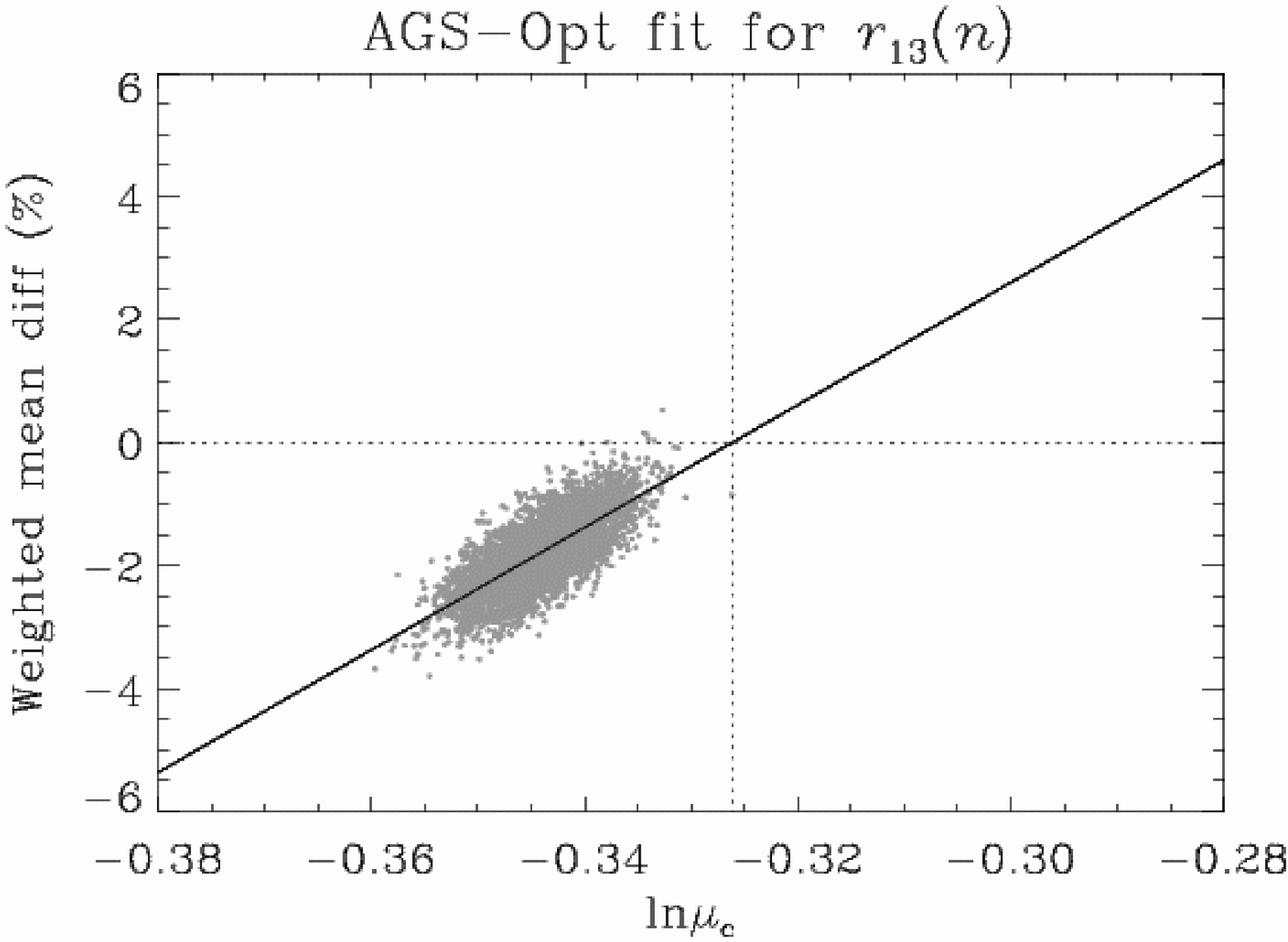}

 \caption{Upper two panels: weighted mean differences, $\langle \Delta
r_{02} \rangle$ (upper left-hand panel) and $\langle \Delta r_{13}
\rangle$ (upper right-hand panel), versus $\ln\mu_{\rm c}$, plotted
for the 1000 Monte Carlo `AGS-Cons' solar models. These models were
made with the `conservative' (large) abundance uncertainties, centered
on the AGS05 mixture. The solid line in each panel is the best-fitting
straight line. The dotted lines intersect at the location along each
best-fitting line where the weighted mean difference is zero. Lower
panels: As per the upper panels, but for the 5000 Monte Carlo
`AGS-Opt' solar models. These models were made with the `optimistic'
(small) abundance uncertainties, centered on the AGS05
mixture. Although visually the solid lines on the lower plots do not
look like the best-fitting lines, they are the unbiased lines given by
a least-squares fit.}

 \label{fig:AGSmuMC}
 \end{figure*}

\begin{figure*}
 \epsscale{1.0}
 \plottwo{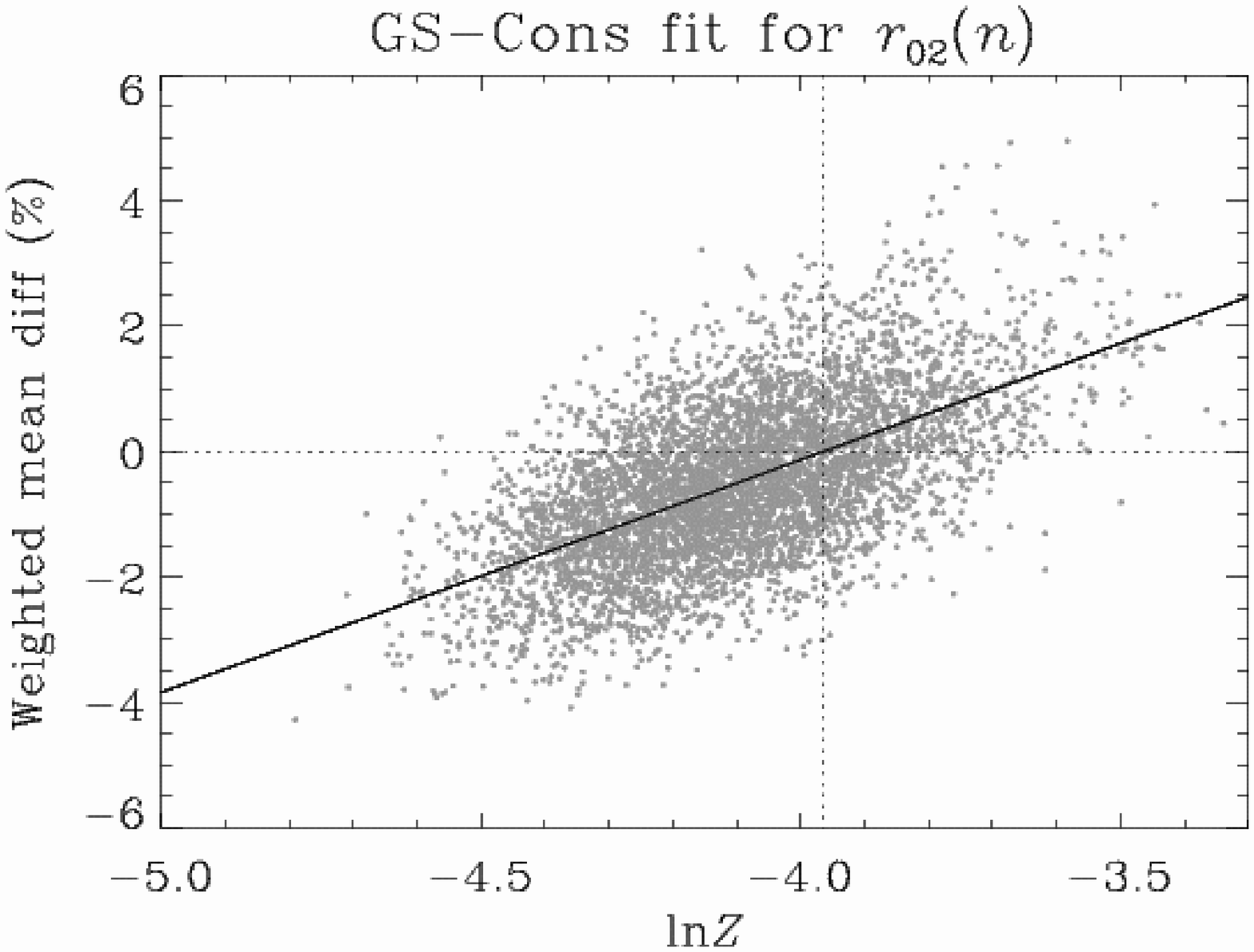}{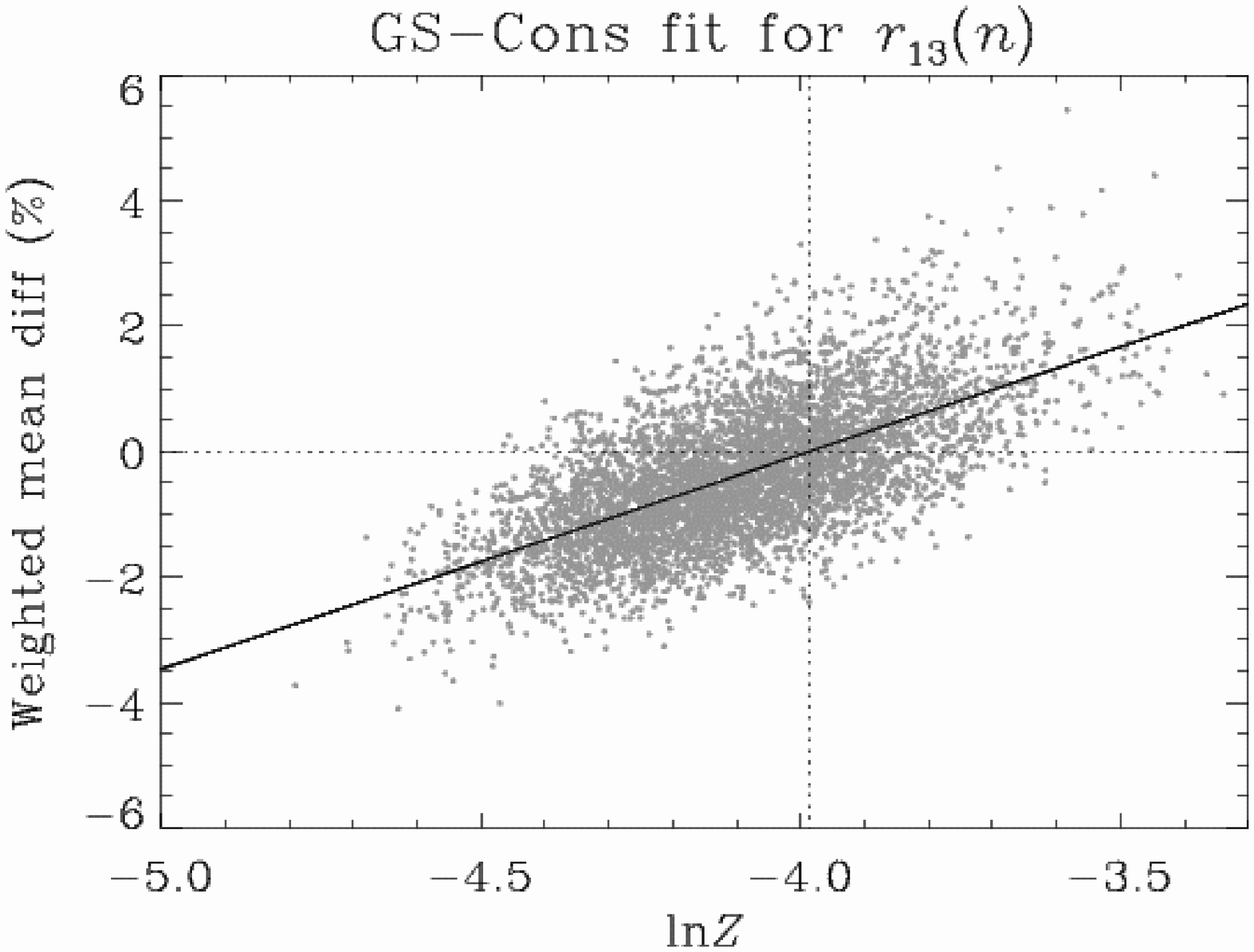}\\
 \plottwo{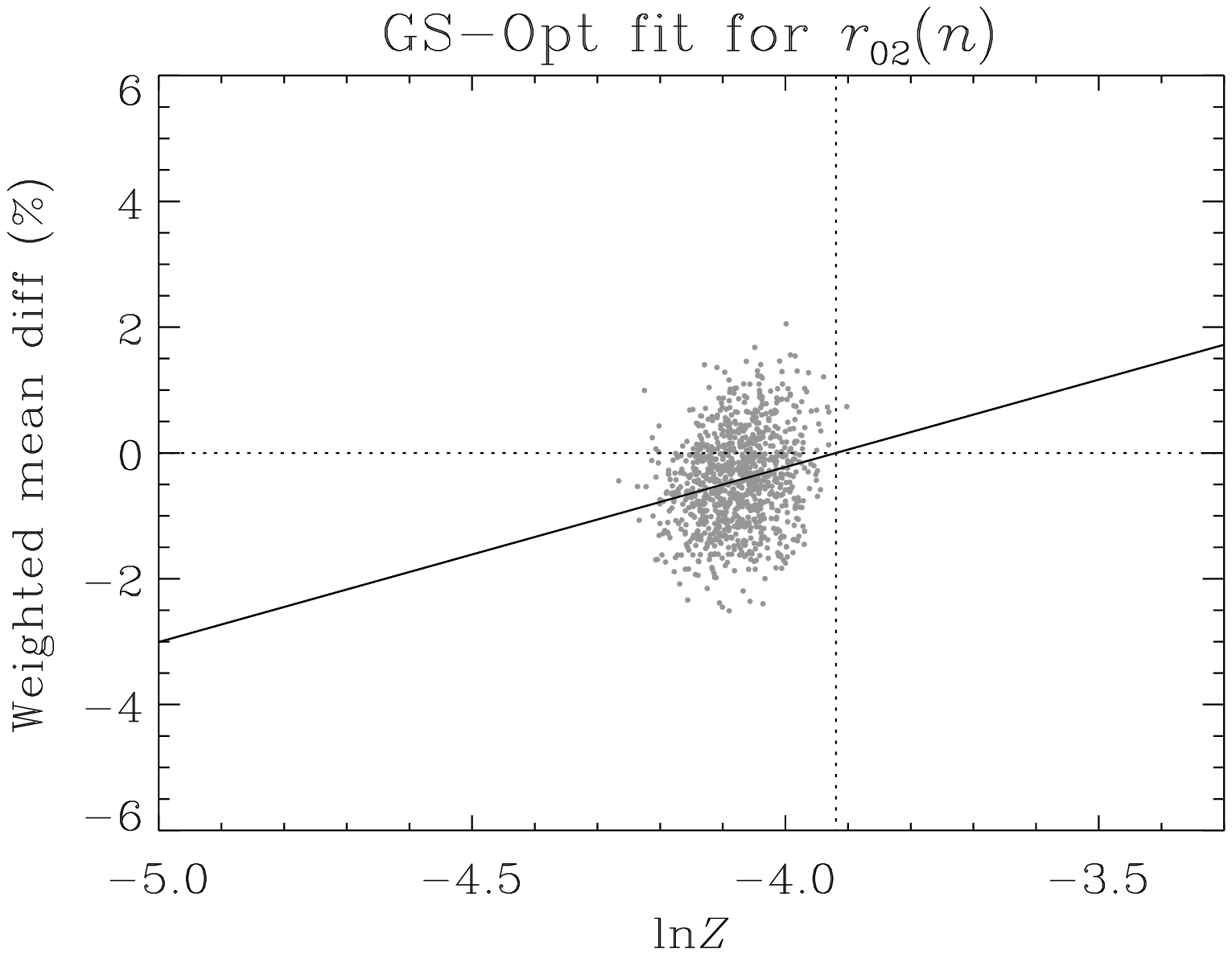}{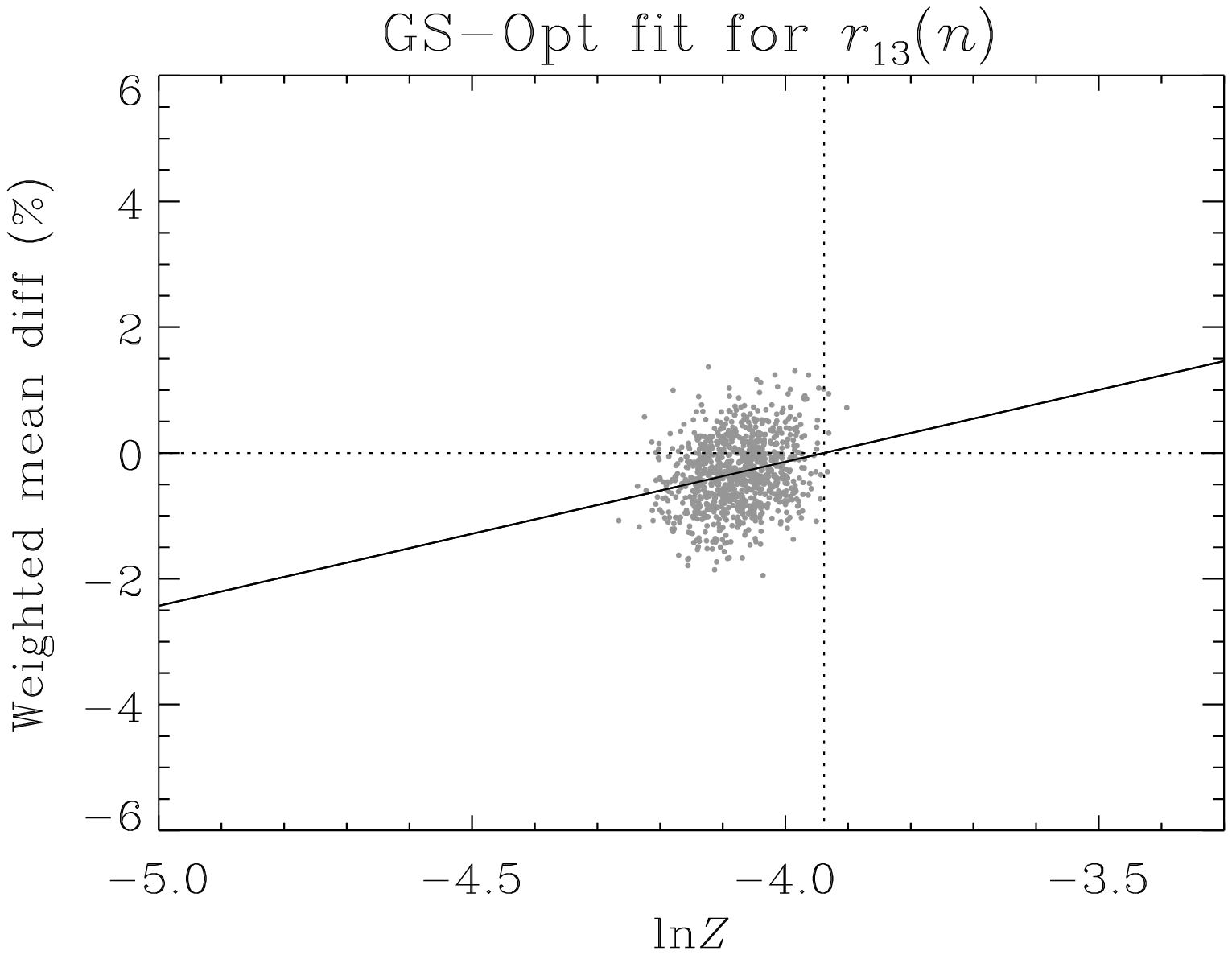}

 \caption{Upper two panels: weighted mean differences, $\langle \Delta
r_{02} \rangle$ (upper left-hand panel) and $\langle \Delta r_{13}
\rangle$ (upper right-hand panel), versus $\ln Z$, plotted for the
5000 Monte Carlo `GS-Cons' solar models. These models were made with
the `conservative' (large) abundance uncertainties, centered on the
GS98 mixture. The solid line in each panel is the best-fitting
straight line. The dotted lines intersect at the location along each
best-fitting line where the weighted mean difference is zero. Lower
panels: As per the upper panels, but for the 1000 Monte Carlo `GS-Opt'
solar models. These models were made with the `optimistic' (small)
abundance uncertainties, centered on the GS98 mixture.}

 \label{fig:GSMC}
 \end{figure*}

\begin{figure*}
 \epsscale{1.0}
 \plottwo{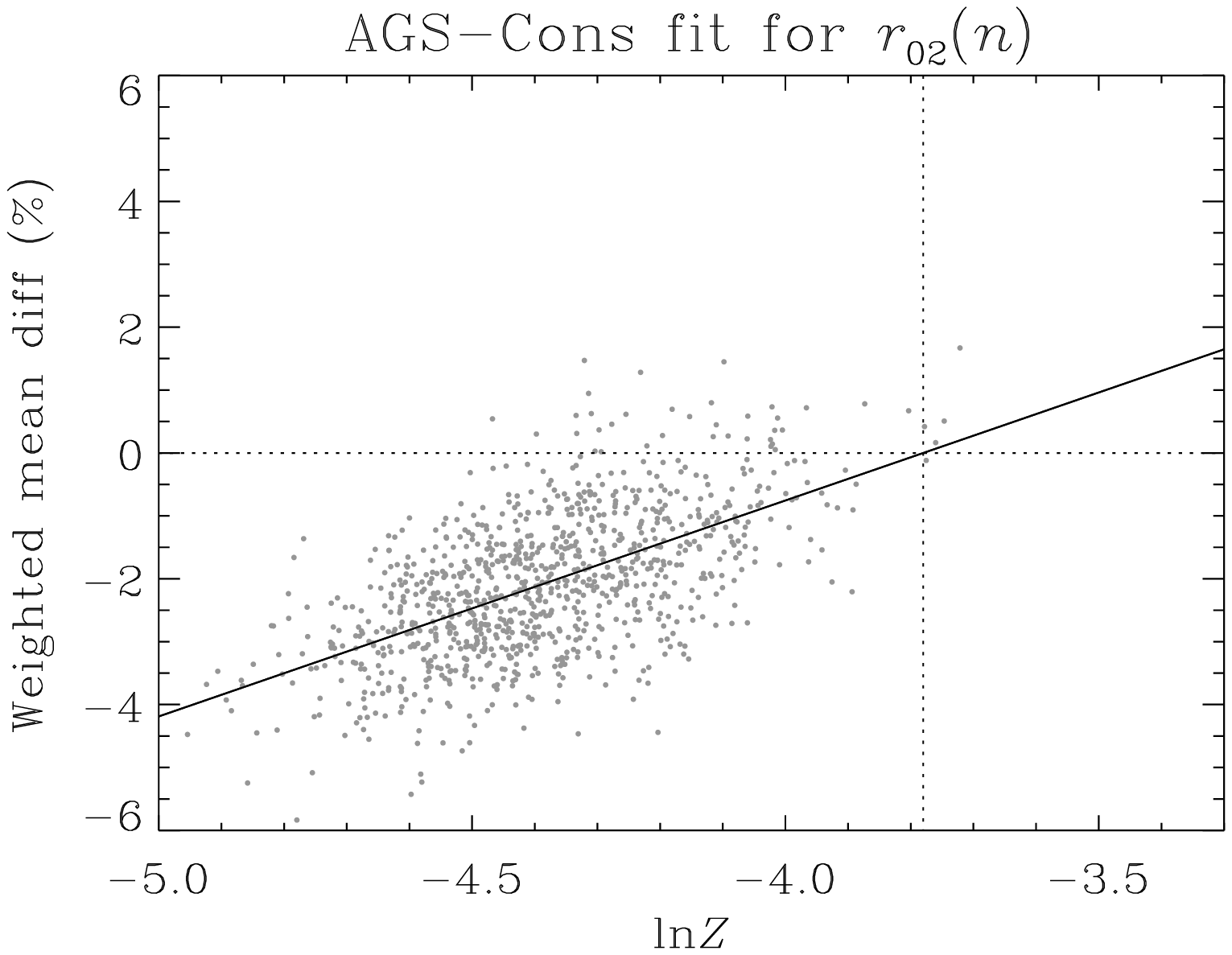}{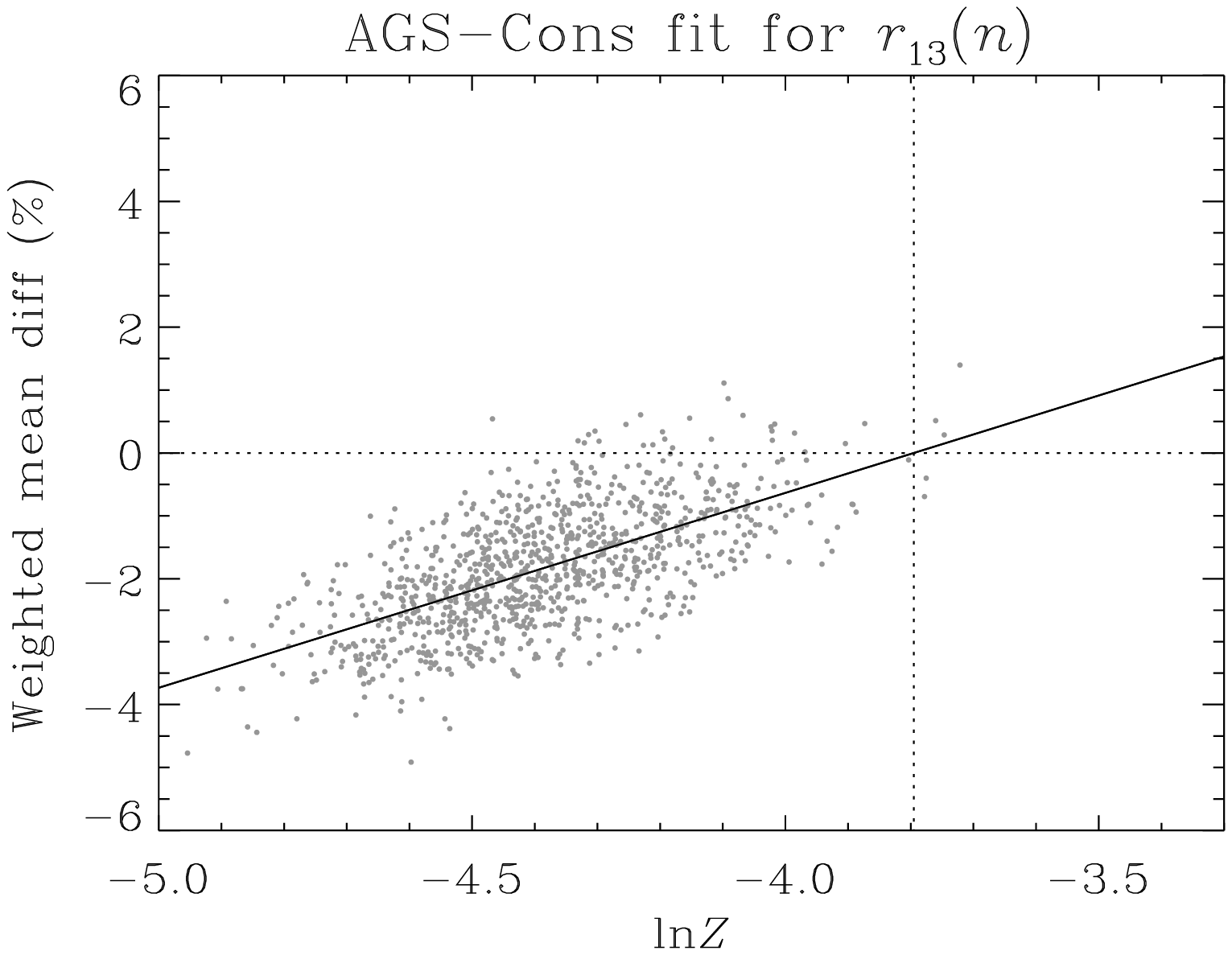}\\
 \plottwo{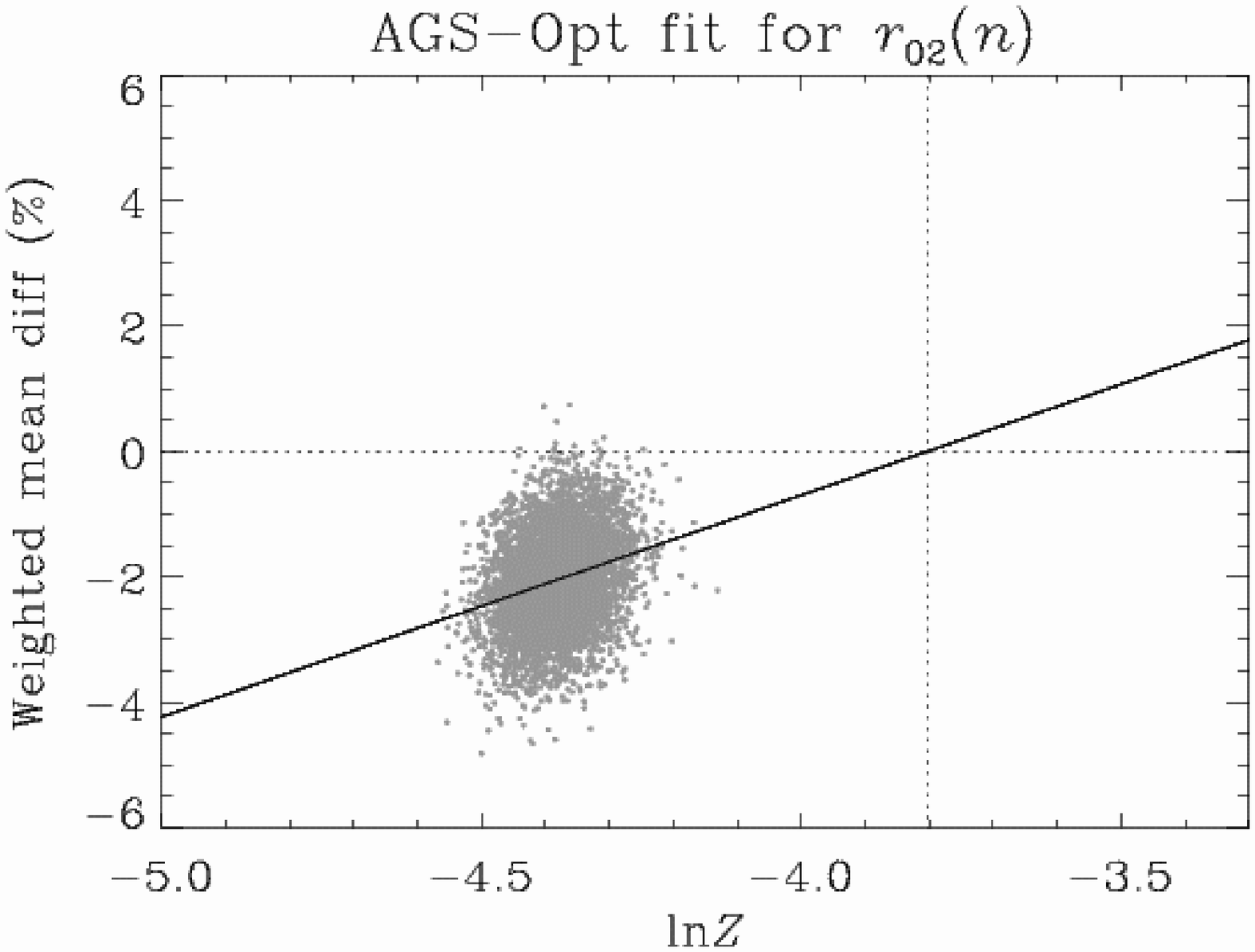}{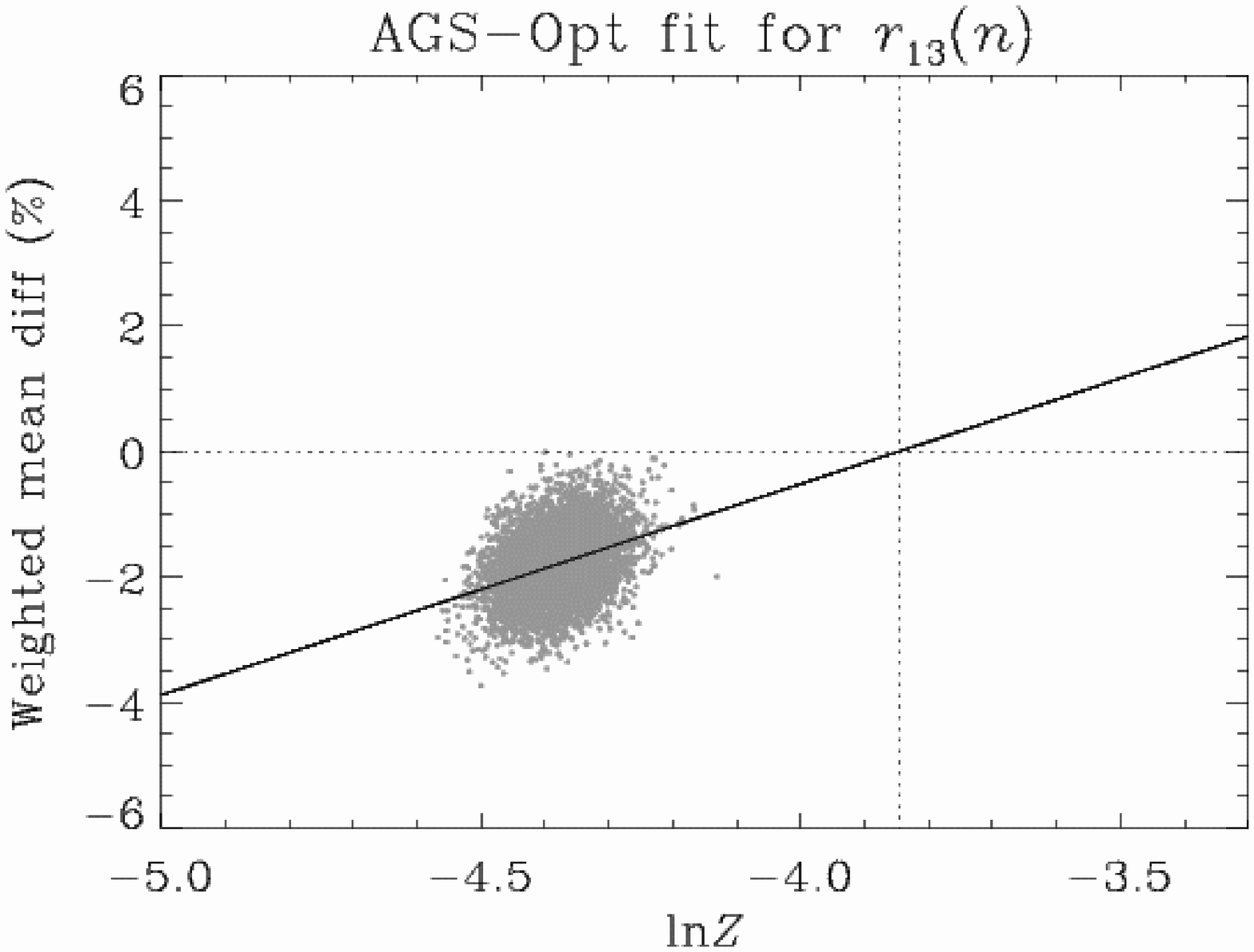}

 \caption{Upper two panels: weighted mean differences, $\langle \Delta
r_{02} \rangle$ (upper left-hand panel) and $\langle \Delta r_{13}
\rangle$ (lower right-hand panel), versus $\ln Z$, plotted for the
1000 Monte Carlo `AGS-Cons' solar models. These models were made with
the `conservative' (large) abundance uncertainties, centered on the
AGS05 mixture. The solid line in each panel is the best-fitting
straight line. The dotted lines intersect at the location along each
best-fitting line where the weighted mean difference is zero. Lower
panels: As per the upper panels, but for the 5000 Monte Carlo
`AGS-Opt' solar models. These models were made with the `optimistic'
(small) abundance uncertainties, centered on the AGS05
mixture. Although visually the solid lines on the lower plots do not
look like the best-fitting lines, they are the unbiased lines given by
a least-squares fit.}

 \label{fig:AGSMC}
 \end{figure*}

Scatter plots of the weighted mean differences of the separation
ratios (in \%) versus the natural logarithm of $\mu_{\rm c}$ are shown
in Figures~\ref{fig:GSmuMC} and \ref{fig:AGSmuMC} for all four sequences
of Monte Carlo models.  Figures~\ref{fig:GSMC} and~\ref{fig:AGSMC} show
the corresponding plots against $\ln Z$. Uncertainties caused by
uncertainties in the radiative opacities and EOS have been included by
adding to the data random components with normal distributions
characterized by the standard deviations given in
Table~\ref{tab:opac_eos}. The solid lines in each panel are the
best-fitting straight lines for the data. The dotted lines intersect
at the location along each best-fitting line where the weighted mean
difference is zero.

We again adopted a linear model for the analysis in the light of: (i)
the results on the test models (discussed in Section~\ref{sec:lin}
above); and (ii) the observed scatter on the plots, which precluded us
from imposing a more complicated fitting model. Detailed breakdowns of
fitting results for the Monte Carlo sequences are presented in
Table~\ref{tab:mcmu} (for $\mu_{\rm c})$, and Table~\ref{tab:mc} (for
$Z$).

The main results obtained from the Monte Carlo sequences are as
follows:

\begin{enumerate}

\item Before turning to discussion of the results on $\mu_{\rm c}$ and
$Z$, we consider first the $\left< \Delta r_{02}\right>$ and $\left<
\Delta r_{13}\right>$ of the model sequences. In the case of the
AGS-Opt sequence, the distribution of $\left< \Delta r_{02}\right>$ is
characterized by a mean value of $-2.05\,\%$ and a standard deviation
of 0.75\,\%.  For $\left< \Delta r_{13}\right>$ the corresponding
values are, respectively, $-1.80\,\%$ and 0.54\,\%.  These values
imply a difference with helioseismology measurements of $2.7\sigma$
and $3.3\sigma$ for $\left< \Delta r_{02}\right>$ and $\left< \Delta
r_{13}\right>$ respectively.  Note from Table~\ref{tab:opac_eos} that
the EOS has a large impact on the total standard deviations of the
$\left< \Delta r_{02}\right>$ and $\left< \Delta r_{13}\right>$
distributions.  As discussed previously, we are probably
overestimating the EOS uncertainties in the separation ratios.
Consequently, the 2.7 and 3.3-$\sigma$ differences should be
considered as robust upper limits to the real discrepancy.

On the other hand, for the GS-Opt sequence the mean and standard
deviation of the $\left< \Delta r_{02}\right>$ distribution are
$-0.43\,\%$ and 0.76\,\% respectively, while for $\left< \Delta
r_{13}\right>$ we get $-0.32\,\%$ and 0.53\,\% for the mean and the
standard deviation.  These numbers translate into differences of only
$\approx 0.6\sigma$ with the helioseismology measurements. It is worth
mentioning that we perform this comparison only for Monte Carlo
sequences having the optimistic choice of uncertainties, because the
aim is to compare helioseismology measurements with solar models that
adopt compositions (central values and uncertainties) given by the
solar abundance determinations, i.e. GS98 and AGS05.

\item As in the case of results obtained using the test models, the
average mean molecular weight in the inner 20\,\% by radius, $\mu_{\rm
c}$ is determined to much higher precision than $Z$ (results in
Table~\ref{tab:mcmu}). The precision in each of the combined measures
of $\mu_{\rm c}$ is better than 0.5\,\%. Even with the improved
precision, the estimates given by analysis of the four sequences of
models are in excellent agreement with one another and provide a very
robust determination of $\mu_{\rm c}$.  The largest difference between any
of the two combined measures is significant at only $\sim
1\sigma$. The average of our four combined seismic estimates is
$\mu_{\rm c}=0.7226$.

\item From the AGS-Opt sequence of models, i.e.  that adopting the
AGS05 central values and uncertainties for the composition, we find
that solar models constructed with the AGS05 composition have an
average $\mu_{\rm c}$ of $0.7088 \pm 0.0029$ (1$\sigma$ uncertainty).
This value differs from our combined seismic estimates of solar
$\mu_{\rm c}$, which are given in Table~\ref{tab:mcmu}, by between 3.1
(for the combined AGS-Cons measure) and $3.7\sigma$ (for the combined
AGS-Opt measure) (where $\sigma$ is determined by adding in quadrature
the uncertainties from the observations and the uncertainty in the
theoretical distribution). On the other hand, in the case of models
adopting the GS98 composition, we derive an average value of $\mu_{\rm
c}=0.7203 \pm 0.0029$ for the optimistic (i.e., lower) uncertainties
on the abundance inputs.  All the combined seismic estimates agree
with this value to well within $1\sigma$.

\item All seismic estimates of solar $Z$ are high (see
Table~\ref{tab:mc}).  They lie noticeably above the low Asplund et
al. (2005a) (AGS05) value.  This is clear even from visual inspection
of the plots in Fig.~\ref{fig:AGSMC}, which shows results for models
with heavy-element mixtures based on the AGS05 values. The location on
the abscissa where the best-fitting lines pass through zero lie well
to the high-$Z$ side of each cloud of points.  None of the AGS-Opt
models has a mean weighted difference greater than zero. Estimation of
$Z$ from these data therefore amounts to an \emph{extrapolation},
rather than the \emph{interpolation} as is possible with the GS-Cons
and GS-Opt models.

From the AGS-Opt sequence of models, which has abundance uncertainties
consistent with Asplund et al. (2005a), we find that solar models
constructed with the AGS05 composition have an average $Z$ of $0.0125
\pm 0.0007$ (1$\sigma$ uncertainty).  This value differs from our
combined seismic estimates of solar $Z$, which are given in
Table~\ref{tab:mc}, by between $2.0\sigma$ (for the combined GS-Cons
measure) and $4.3\sigma$ (for the combined AGS-Opt measure) (where
$\sigma$ is determined by adding in quadrature the uncertainties from
the observations and the uncertainty in the theoretical distribution).

\item All seismic estimates of solar $Z$ are slightly higher than the
Grevesse \& Sauval (1998) (GS98) value, of $Z=0.0169$ (see
Table~\ref{tab:mc}).  This is clear from visual inspection of the
plots in Fig.~\ref{fig:GSMC}, which shows results for models with
heavy-element mixtures based on the GS values. The best-fitting lines
pass through zero in the high-$Z$ parts of the distributions of
points. Differences between the GS98 value and our combined seismic
estimates of solar $Z$ lie between $0.5\sigma$ (for the combined
GS-Cons measure) and $2.0\sigma$ (for the combined AGS-Opt measure).

\item All seismic estimates of solar $Z$ in Table~\ref{tab:mc} are in
good agreement with the seismic estimate of Antia \& Basu (2006),
which was $Z = 0.0172 \pm 0.0020$. The Antia \& Basu result was quoted
with a precision of just over 12\,\%. The precision in our four,
combined measures of $Z$ (rows~3, 6, 9 and 12 of Table~\ref{tab:mc})
ranges from $\sim 12\,\%$ to $\sim 19\,\%$. The largest difference
between any of the two combined measures is significant at only $\sim
1\sigma$.

\end{enumerate}

The observed scatter in Figures~\ref{fig:GSmuMC} to \ref{fig:AGSMC}
deserves discussion, as do the differences between the results given
by the Monte Carlo sequence of models and the test models from
\S~\ref{sec:lin}.

The scatter in the computed $\langle \Delta r_{02} \rangle$ and
$\langle \Delta r_{13} \rangle$ about the best-fitting straight lines
-- as characterized by the $\sigma_{\langle \Delta r_{l,l+2} \rangle}$
in column~3 of Tables~\ref{tab:mcmu} and~\ref{tab:mc} -- far exceeds
the sizes of the formal uncertainties on the mean differences found in
\S~\ref{sec:lin}.  Recall that the formal uncertainty (a result of
uncertainties in the BiSON data) is only $0.046\,\%$ on each value of
$\langle \Delta r_{02} \rangle$; and $0.038\,\%$ on each value of
$\langle \Delta r_{13} \rangle$.  The uncertainties on the seismic
estimates of solar $Z$ and $\mu_{\rm c}$ computed from analysis of the
Monte Carlo data are therefore dominated by scatter introduced by the
input parameter choices for the solar models. As we discuss below, the
largest contribution to this scatter comes from the relative abundance
of heavy elements that contribute to $Z$.  It is worth pointing out
that while the uncertainties on the seismically estimated values of
solar $Z$, obtained with the Monte Carlo sequences, are much larger
(by over an order of magnitude) than the uncertainties obtained from
the test models, corresponding differences in the uncertainties on the
seismic values of $\mu_{\rm c}$, i.e., $\sigma(\mu_{\rm c})$, are just
a few times larger.

The $1\sigma$ standard deviation of the fitting residuals,
$\sigma_{\langle \Delta r_{l,l+2} \rangle}$, changes depending on
whether `conservative' (large) or `optimistic' (small) uncertainties
are used for the input abundances to the solar models.  The
$\sigma_{\langle \Delta r_{l,l+2} \rangle}$ values for the GS-Cons and
AGS-Cons sequences are between $\sim 20$ and $\sim 60\,\%$ higher than
the corresponding values of the GS-Opt and AGS-Opt sequences.  This is
not surprising --- the `conservative' GS-Cons and AGS-Cons sequences
cover, by their very nature, a large, `pessimistic' range of input
abundance values, thus the scatter due to the uncertainties in the
relative abundance is also larger than that for the `optimistic'
sets. The uncertainties that arise from using the GS-Cons and the
AGS-Cons models may therefore be treated as respectable \emph{upper}
bounds for $\sigma(Z)$ and $\sigma(\mu_{\rm c})$.

The gradients and intercepts of the best-fitting lines of the
relations between the separation ratios and the metallicity of the
test models (Figure~\ref{fig:lin}) are steeper and higher,
respectively, then those of their Monte-Carlo counterparts
(Figures~\ref{fig:GSMC}~and~\ref{fig:AGSMC}). The gradients are in
some cases steeper by as much as a factor of 3.  There are also
differences between one Monte Carlo sequence and another, with the
gradients differing by up to $\sim 35\,\%$; the gradients are
constrained typically to much higher precision. However, in spite of
these differences, the zero-crossing points, which serve to provide
the estimates of solar $Z$, return consistently robust values.

Again, it is not very difficult to explain why the gradients are
higher for the sets of test models compared to the sets of Monte Carlo
models, and why the gradients differ from one Monte Carlo set to
another.  The answer again lies in the impact of changes to the
relative mixture of heavy elements.  Each set of linear test models
has the same relative mixture of elements; these models represent the
case of perfect correlation between all the abundance uncertainties. A
given value of $Z$ gives a unique value of $\mu_{\rm c}$, and other
conditions in the core (e.g., temperature), and thus matches to a
unique value of $\langle \Delta r_{02} \rangle$ and $\langle \Delta
r_{13} \rangle$.  The Monte Carlo sequences simulate the opposite
situation; here, all the abundance uncertainties are assumed to be
independent of one another. The different dependences on relative
abundances of $Z$ and $\mu_{\rm c}$ imply that their variations in the
MC sequences become, to some degree, uncorrelated.  For example, large
changes in CNO elements give rise to large changes in $Z$ but have a
much more modest impact on $\mu_{\rm c}$ (and consequently on the
separation ratios). The result is that in the MC sequences the
one-to-one relation between $Z$, $\mu_{\rm c}$ and other conditions in
the core is lost, as is the one-to-one correlation between $Z$ and
$\langle \Delta r_{02} \rangle$ or $\langle \Delta r_{13}
\rangle$. The overall effect is to force the linear fitting gradients
to shallower values when a range of mixtures is admitted in the
models.  There are changes in the details of the mixtures between the
different Monte Carlo sequences, and hence some (albeit much more
modest) changes in gradient are seen from one sequence to another.

It is worth adding that changes to the diffusion rates do not alter
significantly the fitting gradients.  If one takes appropriate subsets
of the Monte Carlo sequences, one finds that the relation between the
rate of diffusion and $\mu_{\rm c}$ is always the same. And the
diffusion rate and $Z$ appear to be only very weakly correlated.
Changes to the diffusion do not affect the relation between the
fitting gradients of $\langle \Delta r_{02} \rangle$ or $\langle
\Delta r_{13} \rangle$ versus $\ln Z$ or $\ln\mu_{\rm c}$. The only
effects seen are changes to the intercept of the fits between the
separation ratios and the metallicities, which change the seismic
estimates of $Z$, and thus diffusion is a relevant source of
uncertainty. We find that physical inputs other than the relative
mixture, the diffusion coefficients, and the EOS have a much smaller
effect on the uncertainties of the estimated values of solar $\mu_{\rm
c}$ and $Z$.

\section{Results and Discussion}
\label{sec:summ}

We have used the frequencies of low-degree acoustic oscillations of
the Sun, determined by the BiSON network over a period of 4752 days,
to try and determine the metallicity $Z$ of the Sun. We did so by
comparing the frequency separarion ratios of a large set of solar
models, from four different Monte Carlo sequences, with the BiSON
observations.  Specifically, we used a weighted average of the
difference of the separation ratios. We find that in addition to giving
good constraints on the solar metallicity, which by definition is the
abundance of heavy elements in the solar convection zone, the
comparison provides an excellent means to determine the mean molecular
weight of the solar core.

The frequency separation ratios of the low-degree acoustic modes are
sensitive to the conditions in the solar core.  By using the weighted
mean differences of the separation ratios, we have obtained seismic
estimates for the mean molecular weight of the solar core ($\mu_{\rm
c}$).  The seismic estimates are robust and depend only very weakly on
the solar models used to construct the weighted mean differences. All
of the four sequences of Monte Carlo models used in this work allow us
to determine $\mu_{\rm c}$ to a precision of better than 0.5\%, and
estimates from different sequences are consistent with each other to
better than $1\sigma$.  This is true even for the seismic estimates
obtained using solar models with the AGS05 composition. These solar
models do not agree with helioseismic results on the solar sound-speed
and density profiles, surface helium abundance, depth of the
convective envelope and frequency separation ratios.  We have now
shown that solar models constructed with the AGS05 composition
(central values and uncertainties) fail to reproduce the seismically
determined average mean molecular weight of the solar core by more
than $3\sigma$.  On the other hand, solar models with the older GS98
composition have $\mu_{\rm c}$ well within $1\sigma$ of our seismic
estimate of $\mu_{\rm c}$ in the Sun.

We get estimates for the solar metallicity in the range between
$Z=0.0187$ and $Z=0.0229$ for the four Monte Carlo sequences with
uncertainties in the range of 12\% to 19\% on each measurement.  All
our seismic estimates for metallicity are consistent with each other
at about the $1\sigma$ level, and are higher than the solar
metallicity derived by Asplund et al. (2005) by between 2.1 and
$4.3\sigma$. Our estimates are also consistent with the seismically
derived value of $Z=0.0172\pm 0.0020$ obtained by Antia \& Basu
(2006). Finally, our estimates are slightly higher (by between 0.5 and
$2\sigma$) than the solar metallicity value recommended by Grevesse \&
Sauval (1998).

The Antia \& Basu (2006) results were obtained by looking at the
near-surface ionization zones. The signature of interest was the
change in the adiabatic index resulting from ionization of material,
which depends on the equation of state and the metallicity. The
biggest source of systematic error in the results of Antia \& Basu was
therefore the equation of state. Here, we have obtained very similar
results for solar $Z$ by looking at a region where very different
physical inputs matter. The main source of systematic error for this
study was the relative abundance of the heavy elements.  Thus errors
in the physical inputs used to construct the solar models are not the
reason why we obtain consistency high estimates of solar $Z$. As a
matter of fact, even when we use low-$Z$ models, we still get
estimates of solar $Z$ that are high.

When the AGS05 abundances were published, it was soon noticed that
solar models constructed with those abundances did not agree well with
the helioseismically determined properties of the Sun. In particular,
the most obvious discrepancy was in the position of the
convection-zone base.  It was speculated that this implied that the
physical inputs to the models, particularly the opacities, were
incorrect. The fact that the helium abundance in the convection zone
was also incorrect lead to the belief that the problems are localized
to the outer parts of the Sun.  Based on the results of this paper, we
conclude that this is not the case and that the problems extend to the
core.  It therefore seems unlikely that the origin of the discrepancy
lies in the simplified modeling of the regions close to the tachocline
or in the treatment of convection which is adopted in standard solar
model calculations.  While rotation and its associated mixing are
certainly not modeled in the standard solar models used in this work,
it should be noted that solar models that attempt to account for
rotation (and some degree of rotationally induced mixing) predict a
lower helium abundance in the solar core, which translates into a
lower average mean molecular weight (Palacios et al. 2006). Our
results seem to indicate that this would make the disagreement between
models with the AGS05 composition and helioseismology even worse.

\acknowledgements This paper utilizes data collected by the Birmingham
Solar-Oscillations Network (BiSON), which is funded by the UK Science
Technology and Facilities Council (STFC). We thank the members of the
BiSON team, colleagues at our host institutes, and all others, past
and present, who have been associated with BiSON. GAV acknowledges the
support of STFC.  SB acknowledges partial support from NSF grant
ATM-0348837.  AMS is partially supported through a John Bahcall
Membership and a Ralph E. and Doris M. Hansmann Membership at the IAS.

\newpage


\begin{deluxetable}{lllll}
\tablecolumns{5} \tablewidth{0pc} \tablecaption{Solar-cycle corrected
BiSON frequencies (in $\rm \mu Hz$) used in the paper} \tablehead{
\colhead{$n$}& \colhead{$l=0$}& \colhead{$l=1$}& \colhead{$l=2$}&
\colhead{$l=3$}} \startdata
 8&                     & $1329.629 \pm 0.004$& $1394.682 \pm 0.011$& $1450.986 \pm 0.038$\\
 9& $1407.481 \pm 0.006$& $1472.841 \pm 0.005$& $1535.861 \pm 0.008$& $1591.575 \pm 0.014$\\
10& $1548.343 \pm 0.008$& $1612.717 \pm 0.006$& $1674.527 \pm 0.008$& $1729.109 \pm 0.017$\\
11& $1686.588 \pm 0.012$& $1749.296 \pm 0.008$& $1810.323 \pm 0.012$& $1865.281 \pm 0.019$\\
12& $1822.212 \pm 0.014$& $1885.089 \pm 0.010$& $1945.804 \pm 0.017$& $2001.218 \pm 0.022$\\
13& $1957.432 \pm 0.015$& $2020.798 \pm 0.014$& $2082.105 \pm 0.018$& $2137.781 \pm 0.025$\\
14& $2093.496 \pm 0.016$& $2156.781 \pm 0.017$& $2217.678 \pm 0.022$& $2273.521 \pm 0.031$\\
15& $2228.768 \pm 0.018$& $2291.980 \pm 0.014$& $2352.196 \pm 0.031$& $2407.660 \pm 0.034$\\
16& $2362.767 \pm 0.025$& $2425.587 \pm 0.019$& $2485.856 \pm 0.026$& $2541.677 \pm 0.032$\\
17& $2496.140 \pm 0.022$& $2559.196 \pm 0.022$& $2619.670 \pm 0.025$& $2676.191 \pm 0.031$\\
18& $2629.656 \pm 0.021$& $2693.332 \pm 0.021$& $2754.454 \pm 0.024$& $2811.352 \pm 0.029$\\
19& $2764.128 \pm 0.021$& $2828.097 \pm 0.021$& $2889.545 \pm 0.024$& $2946.981 \pm 0.029$\\
20& $2899.010 \pm 0.019$& $2963.306 \pm 0.020$& $3024.687 \pm 0.024$& $3082.319 \pm 0.035$\\
21& $3033.736 \pm 0.021$& $3098.127 \pm 0.022$& $3159.800 \pm 0.028$& $3217.712 \pm 0.040$\\
22& $3168.612 \pm 0.025$& $3233.147 \pm 0.026$& $3295.087 \pm 0.035$& $3353.387 \pm 0.054$\\
23& $3303.537 \pm 0.030$& $3368.495 \pm 0.031$& $3430.807 \pm 0.047$& $3489.430 \pm 0.071$\\
24& $3439.006 \pm 0.042$& $3504.195 \pm 0.040$& $3567.005 \pm 0.057$& $3626.022 \pm 0.101$\\
25& $3574.896 \pm 0.055$& $3640.347 \pm 0.052$&                     &                     \\
26& $3710.942 \pm 0.089$&                     &                     &                     \\
  &                     &                     &                     &                     \\            
\hline \enddata
\label{tab:bison}
\end{deluxetable}


\begin{deluxetable}{lllll}

\tablecolumns{5} \tablewidth{0pc} \tablecaption{Model frequencies (in
$\rm \mu Hz$) for exact GS98 abundance} \tablehead{ \colhead{$n$}&
\colhead{$l=0$}& \colhead{$l=1$}& \colhead{$l=2$}& \colhead{$l=3$}}
\startdata
 8&         & 1329.356& 1394.304& 1450.684\\
 9& 1407.157& 1472.596& 1535.579& 1591.042\\
10& 1548.061& 1612.210& 1674.187& 1728.762\\
11& 1686.350& 1748.958& 1809.846& 1864.909\\
12& 1821.823& 1884.864& 1945.720& 2001.096\\
13& 1957.418& 2020.779& 2082.327& 2138.317\\
14& 2093.866& 2157.501& 2218.607& 2274.732\\
15& 2229.760& 2293.493& 2354.395& 2410.169\\
16& 2365.099& 2428.289& 2489.179& 2545.531\\
17& 2499.604& 2563.249& 2624.244& 2681.130\\
18& 2634.334& 2698.557& 2760.298& 2817.711\\
19& 2770.118& 2834.665& 2896.607& 2954.613\\
20& 2906.128& 2971.186& 3033.263& 3091.349\\
21& 3042.414& 3107.438& 3169.828& 3228.318\\
22& 3178.697& 3243.930& 3306.347& 3365.269\\
23& 3314.908& 3380.526& 3443.372& 3502.561\\
24& 3451.650& 3517.369& 3580.551& 3640.271\\
25& 3588.594& 3654.679&         &         \\
26& 3725.807&         &         &         \\
  &         &         &         &         \\
\hline 
\enddata
\label{tab:gs98}
\end{deluxetable}


\begin{deluxetable}{lllll}
\tablecolumns{5} \tablewidth{0pc} \tablecaption{Model frequencies (in
$\rm \mu Hz$) for exact AGS05 abundance} \tablehead{ \colhead{$n$}&
\colhead{$l=0$}& \colhead{$l=1$}& \colhead{$l=2$}& \colhead{$l=3$}}
\startdata
 8&         & 1328.009& 1392.406& 1448.604\\
 9& 1405.414& 1470.904& 1533.686& 1588.631\\
10& 1546.296& 1609.957& 1671.676& 1726.210\\
11& 1684.076& 1746.706& 1807.136& 1861.847\\
12& 1819.239& 1882.140& 1943.031& 1998.140\\
13& 1954.946& 2018.007& 2079.174& 2135.089\\
14& 2090.926& 2154.618& 2215.457& 2271.133\\
15& 2226.755& 2290.126& 2350.930& 2406.628\\
16& 2361.873& 2425.004& 2485.506& 2541.695\\
17& 2496.103& 2559.770& 2620.763& 2677.364\\
18& 2631.043& 2695.017& 2756.563& 2813.952\\
19& 2766.599& 2831.216& 2892.902& 2950.568\\
20& 2902.576& 2967.404& 3029.432& 3087.386\\
21& 3038.792& 3103.708& 3165.737& 3224.098\\
22& 3174.772& 3240.032& 3302.407& 3360.995\\
23& 3311.132& 3376.466& 3439.208& 3498.371\\
24& 3447.692& 3513.433& 3576.329& 3635.823\\
25& 3584.526& 3650.497&         &         \\
26& 3721.747&         &         &         \\
  &         &         &         &         \\
\hline 
\enddata
\label{tab:ags05}
\end{deluxetable}


\begin{deluxetable}{lcccccc}
\tablecolumns{7} \tablewidth{0pc} \tablecaption{$\mu_{\rm c}$
results for `linear test' models} \tablehead{ \colhead{Set}&
\colhead{$N$}& \colhead{$\sigma_{\langle \Delta r_{l,l+2}
\rangle}$}& \colhead{$\mu_{\rm c}$}& \colhead{$\sigma(\mu_{\rm
c})$}& \colhead{$\sigma(\mu_{\rm c})/\mu_{\rm c}$}&
\colhead{$\chi^2_{N-2}$}\\
\colhead{}& \colhead{}& \colhead{(\%)}& \colhead{}& \colhead{}&
\colhead{(\%)}& \colhead{}} \startdata
  GS-Lin $r_{02}$& 10& 0.136& 0.7253& $\pm 0.0012$& 0.17& 9.9\\
 & & & & & \\
  GS-Lin $r_{13}$& 10& 0.099& 0.7244& $\pm 0.0010$& 0.14& 7.4\\
 & & & & & \\
  GS-Lin comb.& & &0.7248& $\pm 0.0008$& 0.11& \\
 & & & & & \\
 & & & & & \\
 AGS-Lin $r_{02}$& 10& 0.108& 0.7260& $\pm 0.0009$& 0.12& 6.1\\
 & & & & & \\
 AGS-Lin $r_{13}$& 10& 0.089& 0.7255& $\pm 0.0008$& 0.11& 6.1\\
 & & & & & \\
 AGS-Lin comb.& & &0.7258& $\pm 0.0006$& 0.08& \\
 & & & & & \\
\hline
\enddata
\label{tab:linmu}
\end{deluxetable}

\begin{deluxetable}{lcccccc}
\tablecolumns{7} \tablewidth{0pc} \tablecaption{$Z$ results for
`linear test' models} \tablehead{ \colhead{Set}& \colhead{$N$}&
\colhead{$\sigma_{\langle \Delta r_{l,l+2} \rangle}$}&
\colhead{$Z$}& \colhead{$\sigma(Z)$}& \colhead{$\sigma(Z)/Z$}&
\colhead{$\chi^2_{N-2}$}\\
\colhead{}& \colhead{}& \colhead{(\%)}& \colhead{}& \colhead{}&
\colhead{(\%)}& \colhead{}} \startdata
  GS-Lin $r_{02}$& 10& 0.057& 0.01798& $\pm 0.00013$& 0.7& 1.7\\
 & & & & & \\
  GS-Lin $r_{13}$& 10& 0.034& 0.01774& $\pm 0.00008$& 0.5& 0.9\\
 & & & & & \\
  GS-Lin comb.& & &0.01785& $\pm 0.00007$& 0.4& \\
 & & & & & \\
 & & & & & \\
 AGS-Lin $r_{02}$& 10& 0.056& 0.01617& $\pm 0.00011$& 0.7& 1.7\\
 & & & & & \\
 AGS-Lin $r_{13}$& 10& 0.050& 0.01604& $\pm 0.00011$& 0.7& 1.9\\
 & & & & & \\
 AGS-Lin comb.& & &0.01611& $\pm 0.00008$& 0.5& \\
 & & & & & \\
\hline
\enddata
\label{tab:lin}
\end{deluxetable}


\begin{deluxetable}{lcccc}
\tablecolumns{6} \tablewidth{0pc}  \tablecaption{Effective standard deviations
  due to uncertainties in the radiative opacities and equation of state}
   \tablehead{  \colhead{Quantity}  &  \colhead{$\Delta
    r_{02}$} & \colhead{$\Delta r_{13}$} & \colhead{$Z$} & 
\colhead{$\mu_{\rm c}$} \\ 
\colhead{} & \colhead{(\%)} & \colhead{(\%)} & \colhead{(\%)} & \colhead{(\%)} } 
\startdata
$\sigma_{\rm opac}$ & 0.16 & 0.064 & 0.083& 0.077\\ 
$\sigma_{\rm EOS}$ & 0.53 & 0.28 & 0.047& 0.031\\
\enddata
\label{tab:opac_eos}
\end{deluxetable}


\begin{deluxetable}{lccccc}
\tablecolumns{6} \tablewidth{0pc} \tablecaption{$\mu_{\rm c}$
results for `Monte Carlo' models} \tablehead{ \colhead{Set}&
\colhead{$N$}& \colhead{$\sigma_{\langle \Delta r_{l,l+2}
\rangle}$}& \colhead{$\mu_{\rm c}$}& \colhead{$\sigma(\mu_{\rm
c})$}&
\colhead{$\sigma(\mu_{\rm c})/\mu_{\rm c}$}\\
\colhead{}& \colhead{}& \colhead{(\%)}& \colhead{}& \colhead{}&
\colhead{(\%)}} \startdata
  GS-Cons $r_{02}$& 5000& 0.82& 0.7234& $\pm 0.0051$& 0.7\\
 & & & & & \\
  GS-Cons $r_{13}$& 5000& 0.56& 0.7229& $\pm 0.0033$& 0.6\\
 & & & & & \\
  GS-Cons comb.& & & 0.7231& $\pm 0.0030$& 0.4\\
 & & & & & \\
   GS-Opt $r_{02}$& 1000& 0.65& 0.7233& $\pm 0.0049$& 0.7\\
 & & & & & \\
   GS-Opt $r_{13}$& 1000& 0.38& 0.7228& $\pm 0.0030$& 0.4\\
 & & & & & \\
   GS-Opt comb.& & & 0.7230& $\pm 0.0026$& 0.4\\
 & & & & & \\
 AGS-Cons $r_{02}$& 1000& 0.76& 0.7214& $\pm 0.0046$& 0.6\\
 & & & & & \\
 AGS-Cons $r_{13}$& 1000& 0.52& 0.7206& $\pm 0.0034$& 0.5\\
 & & & & & \\
 AGS-Cons comb.& & & 0.7209& $\pm 0.0027$& 0.37\\
 & & & & & \\
 AGS-Opt $r_{02}$& 5000& 0.63& 0.7239& $\pm 0.0047$& 0.6\\
 & & & & & \\
 AGS-Opt $r_{13}$& 5000& 0.37& 0.7225& $\pm 0.0029$& 0.4\\
 & & & & & \\
 AGS-Opt comb.& & & 0.7229& $\pm 0.0025$& 0.3\\
\enddata
\label{tab:mcmu}
\end{deluxetable}


\begin{deluxetable}{lccccc}
\tablecolumns{6} \tablewidth{0pc} \tablecaption{$Z$ results for
`Monte Carlo' models} \tablehead{ \colhead{Set}& \colhead{$N$}&
\colhead{$\sigma_{\langle \Delta r_{l,l+2} \rangle}$}&
\colhead{$Z$}& \colhead{$\sigma(Z)$}&
\colhead{$\sigma(Z)/Z$}\\
\colhead{}& \colhead{}& \colhead{(\%)}& \colhead{}& \colhead{}&
\colhead{(\%)}} \startdata
  GS-Cons $r_{02}$& 5000& 1.03& 0.0189& $^{+0.0059}_{-0.0045}$& 28\\
 & & & & & \\
  GS-Cons $r_{13}$& 5000& 0.83& 0.0186& $^{+0.0052}_{-0.0040}$& 25\\
 & & & & & \\
  GS-Cons comb.& & & 0.0187& $^{+0.0039}_{-0.0030}$& 19 \\
 & & & & & \\
   GS-Opt $r_{02}$& 1000& 0.72& 0.0197& $^{+0.0057}_{-0.0045}$& 26\\
 & & & & & \\
   GS-Opt $r_{13}$& 1000& 0.52& 0.0192& $^{+0.0043}_{-0.0035}$& 21\\
 & & & & & \\
   GS-Opt comb.& & & 0.0194& $^{+0.0035}_{-0.0028}$& 16 \\
 & & & & & \\
 AGS-Cons $r_{02}$& 1000& 0.90& 0.0233& $^{+0.0072}_{-0.0055}$& 27\\
 & & & & & \\
 AGS-Cons $r_{13}$& 1000& 0.75& 0.0226& $^{+0.0062}_{-0.0049}$& 25\\
 & & & & & \\
 AGS-Cons comb.& & & 0.0229& $^{+0.0047}_{-0.0037}$& 18 \\
 & & & & & \\
 AGS-Opt $r_{02}$& 5000& 0.72 & 0.0220& $^{+0.0047}_{-0.0039}$& 20\\
 & & & & & \\
 AGS-Opt $r_{13}$& 5000& 0.50& 0.0219 & $^{+0.0038}_{-0.0032}$& 16\\
 & & & & & \\
 AGS-Opt comb.& & & 0.0219& $^{+0.0029}_{-0.0025}$& 12\\
\enddata
\label{tab:mc}
\end{deluxetable}


\begin{thebibliography}{}

\bibitem[Allende Prieto  et al.(2001)]{all01}Allende Prieto,  C., Lambert,
  D. L., \& Asplund, M.  2001, ApJ, 556, L63

\bibitem[Allende Prieto  et al.(2002)]{all02}Allende Prieto,  C., Lambert,
  D. L., \& Asplund, M. 2002, ApJ, 573, L137

\bibitem[Antia \& Basu(2006)]{hma06} Antia, H.M., Basu, S. 2006, ApJ,
644, 1292

\bibitem[Asplund(2000)]{asp00a}Asplund, M. 2000, A\&A, 359, 755

\bibitem[Asplund  et al.(2004)]{asp04}Asplund,  M., Grevesse,  N., Sauval,
  A. J., Allende Prieto, C., \& Kiselman, D. 2004, A\&A, 417, 751

\bibitem[Asplund   et   al.(2005)]{ags05}   Asplund,   M.,
  Grevesse,  N.,  \&  Sauval, A.  J.  2005a,  in  ASP  Conf. Ser.  336,  Cosmic
  Abundances   as   Records   of   Stellar  Evolution   and   Nucleosynthesis,
  ed. T. G. Barnes III, \& F. N.  Bash (San Francisco: ASP), 25

\bibitem[Asplund et al.(2005)]{asp05c} Asplund, M., Grevesse, N.,
Sauval, A. J., Allende Prieto, C., Blomme, R., 2005b,
A\&A, 431, 693

\bibitem[Asplund  et al.(2005)]{asp05a}Asplund,  M., Grevesse,  N., Sauval,
  A. J., Allende Prieto, C., \& Kiselman, D. 2005c, A\&A, 435, 339

\bibitem[Badnell et al.(2005)]{bad05} Badnell, N. R., Bautista, M. A.,
Butler, K., Delahaye, F., Mendoza, C., Palmeri, P., Zeippen, C. J., \&
Seaton, M. J. 2005, MNRAS, 360, 458

\bibitem[Bahcall et al.(2005a)]{bbps05} Bahcall, J. N., Basu, S.,
Pinsonneault, M. H., \& Serenelli, A. M. 2005a, ApJ, 618, 1049

\bibitem[Bahcall et al(2005b)]{bbs05} Bahcall, J. N., Basu, S.,
Serenelli, A.M. 2005b, 631, 1281

\bibitem[Bahcall \& Pinsonneault(2004)]{bap04} Bahcall, J.N.,
Pinsonneault, M.H. 2004, Phys. Rev. Lett., 92, 121301

\bibitem[Bahcall et al.(1997)]{bah97} Bahcall, J.N., Pinsonneault,
M.H., Basu, S., Christensen-Dalsgaard, J. 1997, Phys. Rev. Lett., 78,
171

\bibitem[Bahcall et al.(2005c)]{bsp05} Bahcall, J. N., Serenelli,
A. M., \& Basu, S. 2005c, ApJ, 621, L85

\bibitem[Bahcall, Serenelli \% Basu (2006)]{bah06} Bahcall, J.N.,
Serenelli, A.M., Basu, S., 2006, ApJSS, 165, 400

\bibitem[Basu \& Antia(2004)]{bA04} Basu, S., \& Antia, H. M. 2004,
ApJ, 606, L85

\bibitem[Basu et al. (2007)]{bas07} Basu, S., Chaplin, W. J., Elsworth, Y.
New, R., Serenelli, A. M., Verner, G. A., 2007, ApJ, 655, 660

\bibitem[Basu et al.(2000)]{basu02} Basu, S., Pinsonneault, M.H.,
Bahcall, J.N. 2000, ApJ, 529, 1084

\bibitem[Chaplin et al.(1996)]{chaplin96} Chaplin~W.~J., Elsworth~Y.,
Howe~R., Isaak~G.~R., McLeod~C.~P., Miller~B.~A., New~R.,
van~der~Raay~H.~B.  \& Wheeler~S.~J., 1996, Sol. Phys., 168, 1

\bibitem[Chaplin et al(1999)]{chaplin99} Chaplin~W.~J., Elsworth~Y.,
Isaak~G.~R., Miller~B.~A., New~R., 1999, \mnras, 308, 424

\bibitem[Chaplin et al.(2005)]{chaplin05} Chaplin~W.~J., Elsworth~Y.,
Miller~B.~A., New~R., Verner~G.~A., 2005, \apj, 635, L105

\bibitem[Christensen-Dalsgaard \& Berthomieu(1991)]{jcd91}
Christensen-Dalsgaard, J., Berthomieu, G. 1991, in Solar Interior and
Atmosphere, eds. A.N. Cox, W.C.Livingston, M.S. Matthews, University
of Arizna Press, Tuscon, p401

\bibitem[Christensen-Dalsgaard et al.(1996)]{jcd96}
Christensen-Dalsgaared, J., D\"appen, W., Ajukov, S.V., et al.  1996,
Science, 272, 1286

\bibitem[Delahaye \& Pinsonneault(2006)]{del06} Delahaye, F., \&
Pinsonneault M. H. 2006, ApJ, 649, 529

\bibitem[Elliott \& Kosovichev(1998)]{el98} Elliott, J.R., Kosovichev,
A.G. 1998, ApJ, 500, L199

\bibitem[Ferguson et al.(2005)]{fer05}Ferguson, J.~W., Alexander,
  D.~R., Allard, F., Barman, T., Bodnarik, J.~G., Hauschildt, P.~H.,
  Heffner-Wong, A., \& Tamanai, A. 2005, ApJ, 623, 585

\bibitem[Grevesse \& Sauval(1998)]{gre98} Grevesse, N., \& Sauval,
A. J.  1998, in Solar composition and its evolution --- from core to
corona, eds., C. Fr\"ohlich, M. C. E. Huber, S. K. Solanki, \& R. von
Steiger, Kluwer, Dordrecht, p. 161

\bibitem[Iglesias \& Rogers(1996)]{opal96} Iglesias, C. A., \& Rogers,
F. J. 1996, \apj, {464}, 943

\bibitem[Morel et al.(1999)]{mo99} Morel, P., Pinchon, B., Provost,
J., Berthomiue, G.  1999, A\&A, 350, 275

\bibitem[Ot\'i Froranes et al.(2005)]{oti05} Ot\'i Floranes~H.,
Christensen-Dalsgaard~J., Thompson~M.~J., 2005, \mnras, 356, 671

\bibitem[Palacios et al. (2006)]{pal06} Palacios, A., Talon, S.,
Turck-Chieze, S., \& Charbonnel, C.  2006, to appear in Proceedings of
the SOHO 18 / GONG 2006 / Workshop (ESA SP-624). ``Beyond the
Spherical Sun: a new era for helio- and asteroseismology''.  7-11
August, 2006.  Sheffield, UK.  Editor: K.Fletcher. astro-ph/0609381.

\bibitem[Rogers(2001)]{ro01} Rogers, F. J. 2001, Contrib. Plasma
Phys., 41, 179

\bibitem[1996]{roge96} Rogers,~F.~J., Swenson,~F.~J., \&
Iglesias,~C.~A. 1996, ApJ, 456, 902

\bibitem[Rogers \& Nayfonov(2002)]{opal02} Rogers, F. J., \& Nayfonov,
A.  2002, \apj, 576, 1064

\bibitem[Roxburgh \& Vorontsov(2003)]{rox03} Roxburgh~I.~W.,
Vorontsov~S.~V., 2003, A\&A, 411, 215

\bibitem[Roxburgh(2005)]{rox05} Roxburgh~I.~W., 2005, A\&A, 434, 665

\bibitem[Seaton (2006)]{seaton06} Seaton, M. J., 2005, MNRAS, 362,
L1

\bibitem[Seaton \& Badnell (2004)]{seaton04} Seaton, M. J., Badnell, N. R.,
2004, MNRAS, 354, 457

\end{thebibliography}
\end{document}